\documentclass[12pt,preprint]{aastex}

\newcommand{\dovc}{d_{\rm ovc}}
\newcommand{\lovc}{l_{\rm ovc}}
\newcommand{\rovc}{r_{\rm ovc}}
\newcommand{\rovcz}{r_{\rm ovc}^0}
\newcommand{\rovcmin}{r_{\rm ovc}^{\rm min}}
\newcommand{\rovcmax}{r_{\rm ovc}^{\rm max}}
\newcommand{\thm}{\theta_{\rm m}}
\newcommand{\phim}{\phi_{\rm m}}
\newcommand{\rpc}{r_{\rm pc}}
\newcommand{\rns}{R_{\rm ns}}
\newcommand{\rlc}{R_{\rm lc}}
\newcommand{\rmax}{r_{\rm max}}

\newcommand{\rhomax}{\rho_{\rm max}}
\newcommand{\zobs}{\zeta_{\rm obs}}
\newcommand{\zoo}{\zeta_{\rm obs}^{\rm orig}}
\newcommand{\zop}{\zeta_{\rm obs}^\prime}
\newcommand{\delp}{\Delta^{\rm peak}}
\newcommand{\acc}{\vec a}
\newcommand{\betmin}{\beta}
\newcommand{\betcor}{\beta_{\rm cor}}
\newcommand{\perc}{P}
\newcommand{\prot}{P_{\rm rot}}

\shorttitle{Polarization in Pulsar Models}
\shortauthors{Dyks, Harding, Rudak}

\begin{document}

\title{Relativistic Effects and Polarization\\ in Three High-Energy Pulsar Models}

\author{J. Dyks\altaffilmark{1}, Alice K. Harding}
\affil{Laboratory for High Energy Astrophysics, 
       NASA/GSFC,
    Greenbelt, MD 20771, USA}
\email{jinx@milkyway.gsfc.nasa.gov, harding@twinkie.gsfc.nasa.gov}

\and

\author{B. Rudak}
\affil{Nicolaus Copernicus Astronomical Center, 87-100 Toru{\'n}, Poland}
\email{bronek@ncac.torun.pl}

\altaffiltext{1}{On leave from Nicolaus Copernicus Astronomical Center,
Toru{\'n}, Poland}

\begin{abstract}
We present the influence of the special relativistic effects 
of aberration and light travel time delay on pulsar
high-energy lightcurves 
and polarization characteristics
predicted by three models: the two-pole caustic model, the outer gap model, 
and the polar cap model.
Position angle curves and degree of polarization are calculated
for the models and compared 
with the optical data on the Crab pulsar.
The relative positions of peaks in gamma-ray and radio lightcurves
are discussed in detail for the models.
We find that the two-pole caustic model can reproduce
qualitatively the optical polarization characteristics
of the Crab pulsar -- fast swings of the position angle and minima
in polarization degree associated with both peaks.
The anticorrelation between the observed flux and the polarization
degree (observed in the optical band also for B0656$+$14)
naturally results from the caustic nature of the peaks which
are produced in the model due to the superposition of radiation from
many different altitudes, ie.~polarized at different angles.
The two-pole caustic model also provides an acceptable interpretation 
of the main features in the Crab's radio profile.
Neither the outer gap model nor the polar cap model are able to 
reproduce the optical polarization data on the Crab.
Although the outer gap model is very successful in reproducing the
relative positions of gamma-ray and radio peaks in pulse profiles,
it can reproduce the high-energy lightcurves only when 
photon emission from regions very close to the light cylinder is included. 
\end{abstract}



\section{Introduction}

This paper focuses on two aspects of theoretical modelling of high-energy
emission from pulsars:
1) the influence of special relativity effects on pulsar
high-energy lightcurves, and 2) on linear polarization properties of the
high-energy radiation.

The special relativity (SR) effects which affect the lightcurves include 
the aberration of photon emission directions and time of flight delays
caused by the finite speed of light $c$.
Their importance for lightcurve shapes was recognized by many pulsar
astrophysicists long ago (eg.~Harding et al.~1978; Massaro et al.~1979)
but it was Morini (1983) who first proved that the SR effects themselves
are able to produce prominent peaks in pulsar lightcurves.
Morini obtained the peaks of caustic\footnote{The SR effects result
directly in caustic effects (piling up photons at the same phase of a
pulse). Therefore, hereafter we will use the terms ``special relativity
effects" and ``caustic effects" as synonyms.}
origin in his version of the polar cap model
because he included emission of photons by electrons propagating
at high-altitudes, where the SR effects are important.
The high altitude emission has been often ignored
by polar cap theorists who have naturally focused on near-surface regions,
where the strongest acceleration takes place and
photons of highest energy are produced
(Sturrock 1971; Ruderman \& Sutherland 1975; Daugherty \& Harding 1982; 
Sturner et al.~1995; Rudak \& Dyks 1999).
As does every contemporary model of pulsars, the polar cap model
faces some difficulties with reproducing pulsar data.
Perhaps the most serious of these problems is the difficulty in
reproducing  
the wide separation between the two peaks commonly
observed in the MeV-GeV lightcurves of the brightest gamma-ray pulsars
(Kanbach 1999; Thompson 2001) without invoking very small inclination
angle. 
We discuss the subject in more detail in
Section \ref{polar}.

In the recent version of the outer gap (OG) model 
(Romani \& Yadigaroglu 1995; Cheng, Ruderman, Zhang (2000), 
hereafter CRZ2000) 
peaks in pulsar
lightcurves are purely due to the caustic effects - a fact first emphasized
by Romani \& Yadigaroglu (1995) (hereafter RY95).
However, an important additional factor which determines lightcurve shape
in the OG model is the geometry of the acceleration region (outer gap).
Although the performance of the OG model in reproducing pulsar lightcurves is
relatively good, it fully relies on including photon emission from
the very vicinity of the light cylinder - where the assumed vacuum 
magnetic field geometry is very questionable.
Moreover, the ``traditional" shape of the outer gap (extending from the null
charge surface up to the light cylinder) has recently been questioned
(Hirotani \& Shibata 2002), and this shape was crucial for getting the
good-looking lightcurves.

In view of the problems faced by the afore-mentioned models,
Dyks \& Rudak (2003) introduced an alternative type of 
caustic model which
assumes roughly uniform emission of photons along the entire length of
all last open magnetic field lines, ie. photon emission in the model
extends all the way
from the pulsar surface up to the vicinity of the light cylinder.
The emission is not restricted to the polar gap or to the outer gap region.
There are good prospects
that the slot gap model (Arons \& Scharlemann 1979; Arons 1983)
especially with the recent revisions by 
Muslimov \& Harding (2003a) will provide a viable physical justification
for the assumptions of the extended caustic model.
Acceleration of electrons in regions outside the traditional outer gap 
(ie.~below the null charge surface) has also been recently 
proposed by Wright (2003)
in his empirical model of radio emission from pulsars.
Given that the two-pole caustic model is extremely simple and robust
in explaining pulsar lightcurves with widely separated double peaks
we consider it 
as a serious alternative to both the outer
gap and the polar cap model and we include it in our study.

So far, polarization of pulsar radiation has been most thoroughly
studied within the polar cap model of coherent radio emission.
A geometrical framework of the model was founded already by
Radhakrishnan \& Cooke (1969). Much later Blaskiewicz et al.~(1991)
furnished the model with lowest order special relativity
effects.
However,
the noncoherent high-energy emission has a different spatial origin
and requires a separate study.
Polarization properties of the high-energy radiation have been 
concisely discussed for both 
the polar cap (Daugherty, Harding 1996)
and the outer gap model (Romani
Yadigaroglu 1995; Chen et al.~1996). We undertake the calculations and
discussion again, because of the complexity of the subject, as well as
because of access to new, excellent quality, optical polarization data on
the Crab pulsar (Kanbach et al.~2003). 
Also, upcoming progress in high-energy polarimetry
(Integral, Mega, Advanced Pair Telescope) 
makes the subject of polarization very important. 

The paper is organized as follows:
Section 2 describes the way in which we calculate lightcurves and
polarization properties in the models described above.
Section 3 presents our results.
In Section 3.1 we discuss
the two-pole caustic model, in Section 3.2 -- the outer gap model, 
and in Section 3.3 -- the polar cap model.
Section 3.4 includes
a discussion of relative positions of gamma-ray and radio peaks 
in pulsar lightcurves. Interpretation of radio pulse profile of the Crab 
pulsar within the two-pole caustic model is given there.
Section 4 contains conclusions.

\section{Calculation method}

We assume that the magnetic field of a pulsar has the geometry of a
vacuum dipole distorted by rotational effects, ie. we use the ``retarded
vacuum dipole" approximation, commonly encountered in previous
research (eg.~Romani \& Yadigaroglu 1995; Arendt \& Eilek 1998;
Cheng et al.~2000).
However, to enable estimate of the importance of near light-cylinder magnetic 
field distortions,
we also present a limited number of results for the static shape dipole.

We perform Runge-Kutta integrations along magnetic field lines
to determine a shape of the polar cap rim.
With the polar cap rim determined, we calculate 
``open volume coordinates" (defined below) at the star surface 
to position footprints
of those magnetic field lines along which photon emission is then
followed. These steps are described in more detail in the following
subsections.
The photon emission is then modelled as follows: 
the field lines in a frame
corotating with the star
are divided into small segments, the length of which had been
constrained by the pre-determined phase resolution of our calculation
(typically $360$ bins per rotation period).
For each segment the position
of emission point $\vec r_{\rm em}^{\ \prime}$ and the direction of 
emission $\vec \eta^{\ \prime}_{\rm em}$ in the (primed) corotating frame
(CF) are determined. The emission direction is assumed to be along the
electron velocity in the CF, ie.~tangent to the magnetic field line. 
Then the Lorentz transformation of 
$\vec \eta^{\ \prime}_{\rm em}$ (aberration) to the (unprimed) 
value in an inertial observer frame (IOF)
is performed. The direction of photon propagation $\vec \eta_{\rm em}$
in the IOF 
along with $\vec r_{\rm em}$ determine the phase $\phi$ 
at which photons are observed. Time delays due to the finite speed of light 
are taken into account at this stage.  
The electric field vector of the
emitted ``wave" $\vec E_{\rm w}$ in IOF is determined in each step as follows:
a new emission direction $\vec \eta_1^{\ \prime}$ 
at a nearby emission point at the same magnetic field line is found, 
$\vec \eta_1^{\ \prime}$ is rotated to account for the spin of the neutron
star,  and finally transformed to the IOF, 
where the difference between $\vec \eta_{\rm em}$ and $\vec \eta_1$
is used to calculate the acceleration
of the electron at the emission point $\acc$.
We followed Blaskiewicz et al.~(1991, hereafter BCW91) 
and Hibschman \& Arons (2001, hereafter HA2001)
by assuming that $\vec E_{\rm w}$ is parallel to $\acc$.
This approach differs from the method of Radhakrishnan \& Cooke (1969) 
and RY95 who assumed that
the vector $\vec E_{\rm w}$ has the direction of the curvature radius
of a magnetic field line at the emission point
which did not include the acceleration due to rotation. 
To emulate uniform
intensity, the Stokes parameter $I$ is simply assumed to be equal   
to the length of a field line segment. 
To calculate the parameters $Q$ and $U$
we use the position angle (PA) $\psi$ between $\vec E_{\rm w}$ and the
projection of the pulsar spin axis on the plane of the sky.
Propagation effects are ignored and
the emitted radiation is assumed to be linearly polarized at 80\%.
The values of $I$, $Q$, and $U$ are then
accumulated in 
(ie.~added to those already present in) appropriate bins
of 2D tables, 
the two dimensions of which correspond
to the phase
$\phi$ and to the observer's position $\zobs = (\vec \eta_{\rm em})_z$.
After collecting photons from all field line segments appropriate for a
given model (eg.~for the outer gap model, only outward emission 
from regions
above the null charge surface is included), the   
final values of Stokes parameters (from the 2D tables) are transformed 
back into values of position angle $\psi=0.5 \arctan{(U/Q)}$ 
and polarization percentage $\perc=(Q^2 + U^2)^{1/2}/I$.

Although the above description may give the impression that we followed exactly
the methods from previous works (eg. RY95; CRZ2000), 
our calculation differs from them in a few important details
which are discussed below.

\subsection{The shape of a polar cap}

In previous studies (RY 95; Arendt, Eilek 1998; CRZ2000)
the polar cap rim was determined via bisection in magnetic colatitude
$\thm$ performed at fixed magnetic azimuths $\phim$, with the latter
spaced uniformly around the magnetic pole.  
This method resulted in a discontinuous rim shape -- with a  
``jump" or a ``glitch"
at $\phim\sim 115^\circ$ for moderate dipole inclinations $\alpha =
40^\circ - 50^\circ$.

In our calculations, we applied bisection in the magnetic azimuth $\phim$ 
(at fixed $\thm$) for the glitch part of the polar cap rim, and the
traditional bisection in $\thm$ (at fixed $\phim$) for the rest of the
rim. 
This produced the interesting result presented with a thick solid line
in Fig.~1 (for $\alpha = 45^\circ$): the ``glitch"
actually appears to be a ``notch" in the polar cap rim.
The rim winds ``backwards" at the notch, where, for a fixed magnetic azimuth
$\phim$, there are \emph{three} solutions for the rim's magnetic colatitude
$\thm$.

The tip of the notch (marked with C1 in Fig.~\ref{caps2}) is one of two
``critical" points on the polar cap rim. The second critical
point is marked by C2 in Fig.~\ref{caps2}. The critical points divide
the polar cap rim into 2 parts, denoted by R1 and R2 in
Fig.~\ref{caps2}. As first described by Yadigaroglu (1997), 
magnetic field lines
emerging from the R1 part of the rim are tangent to the light cylinder
at points forming a spiral line. Lines emerging from the R2 part of the
rim are
tangent to the light cylinder at points forming a different spiral.
In this sense the rim of the polar cap consists of two different curves
(R1, and R2) which meet at the critical points.
As will become clear in Section 3, a high-altitude spread of magnetic
field lines with footprints distributed uniformly along the polar cap
rim, changes discontinuously at the critical points C1 and C2.
This produces notable effects in lightcurves.

Although the electric field near polar gap should destroy (or at least
smooth out) the notch, for definiteness we include it in our calculations.
We find that the high-altitude spatial spread of magnetic field
lines which emerge from the notch part of the polar cap rim 
(on the R1 side of the C1 point) is much larger
than the high-altitude spread 
of lines emerging from any other part of the polar
cap rim, assuming uniform distribution of footprints along the rim.
Therefore,
neglecting the notch part of the polar cap rim 
results in a very large part
of magnetosphere void of magnetic field lines.
For $\alpha = 45^\circ$, this ``missed region" extends over 
more than a half of one quadrant of the magnetosphere
(cf.~fig.~2 in Arendt \& Eilek 1998) .

\subsection{Open volume coordinates}

Rotational effects destroy the symmetry of the dipolar magnetic field around
the dipole axis. Therefore the magnetic pole no longer provides a useful
reference point at the stellar surface --
the primary reference object becomes the rim of polar cap.
This forced Yadigaroglu (1997) and CRZ2000 to introduce ``open volume
coordinates". One of the coordinates was the magnetic azimuth $\phim$ 
of a point at
the star surface and the second one (denoted by $a$ by CRZ2000) 
was equal to the ratio $\theta_{\rm m}/\theta_{\rm m}^{\rm rim}$, where 
$\theta_{\rm m}$ is the magnetic colatitude of that point,
and $\theta_{\rm m}^{\rm rim}$ is the magnetic colatitude of the polar cap rim
measured at $\phim$. 
Yadigaroglu (1997) used a similar coordinate: $w=1 - a$.

Although connected to the polar cap rim, these coordinates still
refer to the magnetic pole and, therefore, have serious disadvantages:
1) they cannot be defined for moderate inclination angles $\alpha =
40^\circ - 50^\circ$ because of the ambiguity of rim colatitude at the
notch;
2) even for cases without the notch (eg.~for $\alpha > 60^\circ$ often
considered in the outer gap model), different points having the same
value of $a$ in general 
do not lie at the same distance from the polar cap rim.   

Therefore, we introduce new open volume coordinates 
$(\rovc, \lovc)$ to identify points at the star surface.
The first coordinate is defined as
$\rovc=1 \pm \dovc$, where $\dovc$ is the minimum distance of a
point from the polar cap rim normalized by the standard polar cap radius
$\rpc = (\Omega\rns^3c^{-1})^{1/2}$;
($\Omega=2\pi/\prot$ is the angular velocity of pulsar rotation, 
$\prot$ is the rotation period, and $\rns$ is
the radius of the neutron star). The plus sign refers to points
lying outside the polar cap, the minus sign is for points within the cap.
Thus, all points with, eg.~$\rovc=0.9$ form a deformed ring
(similar in shape to the polar cap rim) lying inside the polar cap
at fixed distance $0.1\rpc$ from the rim.
For points at the rim $\rovc=1$.
Rings of fixed $\rovc$ are shown in Fig.~2.
For small ($\alpha \la 30^\circ$) and large ($\alpha \ga 65^\circ$)
dipole inclinations,
$a$, $w$, and $\rovc$ are approximately related by:
$\rovc \simeq a$, and $\rovc \simeq 1 - w$.\\
The second coordinate $\lovc$ of a point located at a given
$\rovc$ is the arclength measured along the deformed ring 
of fixed $\rovc$ at which the considered point lies.
The arclength is measured in the direction of increasing $\phim$
(counterclockwise in Fig.~2), 
with $\lovc=0$ corresponding to $\phim=0$.   

Although being difficult to establish, the new open volume coordinates
enable us to easily emulate useful electron distributions at the star
surface; eg.~to model photon emission due to a uniform distribution of
electrons between some $\rovcmin$ and $\rovcmax$
we calculate the equidistant deformed rings within the considered
range of $\rovc$, then we position magnetic field
line footprints uniformly along each ring (ie.~nonuniformly in $\phim$).
Since we follow photon emission for the same number of magnetic field
lines for each ring, the emissivity for different rings is weighted
by $l_{\rm ring}/l_{\rm rim}$, where $l_{\rm ring}$ is the ``circumference"
of a given ring, and $l_{\rm rim}$ is the ``circumference" of the polar
cap rim. 
 
Since most contemporary pulsar models consider the last open magnetic
field lines as a primary region of photon emission (or lines close to
the last open) below we also consider a ``rim dominated" electron
density distribution at the star surface:
we additionally weight
the emissivity from different rings by a value of the Gaussian function:
\begin{equation}
\frac{dN_{\rm ph}}{ds} \propto
\exp\left(-\frac{(\rovc - \rovcz)^2}{2\sigma^2}\right) 
\label{gauss}
\end{equation}
with $\sigma$ usually set equal to $0.025$. 
The function is centered at the polar cap rim ($\rovcz = 1$, two-pole
caustic model) or at $\rovcz = 0.9$ (outer gap model).
Fig.~\ref{inelpos2} presents a sample distribution 
of magnetic field line footprints 
at the star surface for the lines along which photon emission is then
followed.

\subsection{Assumed emissivity}

For the two-pole caustic and for the outer gap model, 
we assume a uniform emissivity along the magnetic field lines
as a useful preliminary approximation of actual photon emission.
This approximation is reasonable since
the electron energy may be stabilized within a very broad range of altitudes
by the radiation reaction limited acceleration 
(eg.~see fig.~10 in Hirotani et al.~2003). 
Without a reliable prescription for an altitude at which the
emission ceases, the ``cut-off" altitude becomes an important,
additional parameter of our calculations.

Since the geometry of the dipolar magnetic field is not known close to
the light cylinder, when modelling emission for the two-pole caustic model 
we assume that the emissivity drops sharply to zero
at a distance $\rhomax$ from the rotational axis, with
$\rhomax$ between $0.75$ and $0.95 \rlc$, where $\rlc = c/\Omega$ is the
radius of the light cylinder.
This distinguishes our calculation from those of RY95, Yadigaroglu
(1997), and CRZ2000, who assumed that the strong
photon emission can be
reliably traced up to $\rhomax = \rlc$.
Given that the geometry of magnetic field lines close to the light
cylinder should be strongly influenced by magnetospheric currents
and particle inertia we do not consider the near-$\rlc$ region
of the vacuum retarded dipole a good approximation of the 
real magnetic field.
To support the vacuum retarded dipole it is often argued that
it approximates the actual magnetosphere at least in the limit
of low particle density, which is expected within vacuum gaps.
However, since the open field lines are assumed 
to be a primary source for the bulk
of gamma-ray emission 
they must be non-negligibly loaded with
charged particles which makes the vacuum dipole approximation
not applicable to them.
Moreover, currents generated by charges which fill in regions 
outside the ``vacuum" gaps (regions adjacent to the gaps in particular)
should modify the magnetic field
not only outside the gaps, but also inside them.\\
In the outer gap model the bulk of gamma rays comes from ``open" field lines
lying \emph{close} to the last open lines -- with $\rovc \simeq 0.9$
(RY95, CRZ2000). 
However, the ``close to last open" field lines in fact \emph{close}
just behind the light cylinder, where they should open to connect smoothly 
with a wind region.
This change in the field line geometry 
would propagate inwards, and would influence the shape of field lines
within, and close to the light cylinder.

Therefore, we argue that reliable outer gap calculations should be
limited to $\rhomax\sim 0.8\rlc$.
However, outer gap pulse profiles calculated for $\rhomax = 0.8\rlc$
do not exhibit the leading peak, which in the outer gap model forms
very close to the light cylinder.
Therefore, to allow for the leading peak to be formed, and to 
enable comparison with previous results 
below we present outer gap results for $\rhomax\approx\rlc$.
 
Apart from the emission boundary due to the proximity of the light cylinder,
it is useful to constrain the photon emission to some limited
distance from the star. This would take into account a probable decline of
emissivity with increasing altitude.
For example, Romani \& Yadigaroglu (1995) assume 
a Gaussian decline in emissivity 
(with $\sigma = 0.5 \rlc$) at distances $s$ measured 
along magnetic field lines larger than $\rlc$.
In this paper we assume a constant radial distance from the star $\rmax$
as an additional upper boundary of the emission region.
For $r > \rmax$ we assume zero emissivity.   

\subsection{The position angle -- conventions}
\label{conventions}

Contrary to BCW91
and HA2001 we assume that
the position angle $\psi$ increases
\emph{counterclockwise} on the sky, ie.~we follow the usual
astronomical convention (Damour \& Taylor 1992; 
Everett \& Weisberg 2001, hereafter EW2001).
HA2001
studied the influence of rotational effects and magnetospheric currents
on the shape of the position angle curve and found that 
some of these effects result in vertical shifts of 
the entire PA curve. 
Since the conventional position angle $\psi$ 
corresponds to the negative of the
position angle $\psi_{\rm HA}$ considered by HA2001 
(ie.~$\psi =-\psi_{\rm HA}$), 
the rotationally-induced vertical shifts of the conventional
PA curve $\psi(\phi)$ occur in the opposite direction
than the shifts of the curve $\psi_{\rm HA}(\phi)$ described by HA2001.
In particular, the aberration-induced decrease (or downward shift)
of $\psi_{\rm HA}$ found by HA2001 for $\zeta <
90^\circ$ corresponds to the upward
shift of the conventional PA curve $\psi(\phi)$.

In figures with model results we present position angle curves
only for viewing angles $\zobs \le 90^\circ$, since the position angle
is antisymmetric with respect to the rotational equator. 
Thus, an observer located at $\zobs^\prime = 180^\circ
- \zobs$ records the PA curve $\psi^\prime(\phi)$
which is a mirror image (with respect to $\psi=0$) of the PA
curve $\psi(\phi)$ recorded by an observer located at $\zobs$.
More precisely,
the mirror image is shifted horizontally in phase by half the
rotation period, ie.~$\psi^\prime(\phi)=-\psi(\phi+\pi)$.

The modelled position angle curves are symmetric with respect to
the transformation $\alpha \rightarrow \pi - \alpha$, because the velocity
of electrons (used to determine $\acc\parallel \vec E_{\rm w}$)
is assumed to be directed outwards in both magnetic 
hemispheres, so that the information about the direction of the
positive dipole axis ($+\vec \mu$) is lost. 
The transformation only shifts the position angle curves 
horizontally by half the rotation period: $\psi^\prime(\phi) = \psi(\phi +
\pi)$. The sign of $\psi$ is not affected (no mirror reflection).
Obviously, unlike in the case of the \emph{modelled} position angle curve,
the horizontal shift by half the rotation period cannot
be recognized in the \emph{observed} PA curve 
(ie.~it does not allow us to
discern between $\alpha$ and $\pi-\alpha$) because the assignment
of the zero phase to the observed PA curve is arbitrary
and the pulse profile undergoes the same shift as the PA curve
(ie.~$I^\prime(\phi) = I(\phi+\pi)$).
The behaviour of the position angle curve under the reflections of 
the dipole moment $\vec \mu$ and the observer's position with respect to the
rotational equator is summarized in Table 1.

\begin{deluxetable}{ccc}
\tabletypesize{\scriptsize}
\tablecaption{Symmetries of the position angle curve with respect to
the transformations $\alpha \rightarrow \pi - \alpha$ and $\zobs \rightarrow
\pi - \zobs$. The third column gives the PA curve 
$\psi^\prime(\phi)$ which
is recorded by an observer located at $\zop$ and who is viewing a pulsar
with the magnetic dipole tilted at the angle $\alpha^\prime$ with respect to
the positive rotational axis, \emph{if} another observer (located at $\zobs$ 
and viewing the dipole tilted at $\alpha$) records
the PA curve $\psi(\phi)$.
\label{table1}}
\tablewidth{0pt}
\tablehead{
\colhead{$\alpha^\prime$} & \colhead{$\zop$}   & 
\colhead{$\psi^\prime(\phi)$}
}
\startdata
$\alpha$ & $\pi - \zobs$ & $-\psi(\phi+\pi)$ \\
 & & \\
$\pi-\alpha$ & $\zobs$ & $\psi(\phi+\pi)$ \\
 & & \\
$\pi-\alpha$ & $\pi-\zobs$ & $-\psi(\phi)$ \\
 \enddata
\end{deluxetable}

Thanks to the symmetries 
presented in Table 1,
a fit of the modelled position angle curve to the high-energy polarization
data enables unambiguous determination of the viewing angle $\zobs$,
but \emph{does not} allow us to discern between $\alpha$ and $\pi-\alpha$.
The value of $\zobs$ usually cannot be uniquely derived from fitting the PA 
curves to \emph{radio} data (eg.~EW2001), because the latter are usually 
limited to a very narrow range of rotational phase.
Pulsar profiles at high photon energies
usually have significantly larger duty cycles than at the radio frequencies,
which should remove the problem of determining $\zobs$ 
as soon as high quality polarization data and reliable models
of position angle curves at high energy bands 
are at hand.\footnote{Our conclusion, that the fit of the position angle
curve makes it possible to
unambiguously determine the value of $\zobs$ but not $\alpha$,
may seem to be contradicted by the wording of EW2001, who claim
that the fit enables the unambiguous determination of ``$\alpha$". 
Actually, however, the apparent disagreement
results from different definitions of $\alpha$.
In this paper we assume that $\alpha$ is the angle between the positive
rotation axis (pointing in the direction of $\vec \Omega$) and the magnetic
moment of the dipole $\vec \mu$
(pointing toward the magnetic north). EW2001 use $\alpha_{\rm EW}$
defined as the angle
between $\vec \Omega$ and the \emph{observable magnetic pole},
regardless of whether the pole is magnetically northern or southern.
Obviously, the value of $\alpha_{\rm EW}$ defined in this way, 
along with the value of
$(\zobs-\alpha_{\rm EW})$ which EW2001 derive from their fits, 
provide unambiguous information
about the location of the observer (ie.~about $\zobs$) rather than about
the direction of $+\vec \mu$. 
The angle $\alpha\equiv\angle(+\vec \Omega, +\vec \mu)$ 
cannot be uniquely determined from the fits of modelled
position angle curves to pulsar data, regardless of the
frequency of the observation (ie.~radio or high-energy).
The fits provide us only with $\sin\alpha$, ie.~do not allow to discern
between $\alpha$ and $\pi-\alpha$.} 

Also because of the symmetries presented in Table 1, the position angle
curves calculated for the clockwise definition of position angle 
(eg.~those
shown in figures in HA2001) can be considered as the position angle
curves for the standard (ie.~counterclockwise) definition of $\psi$,
but for viewing angles $\zobs = \pi - \zoo$ and the dipole
inclinations $\alpha = \pi-\alpha^{\rm orig}$, where
the angles with the superscript `orig' are those given 
in the published works with the clockwise definition of the position
angle (cf.~Table 1, last row).

Finally, we recall that when calculating the modelled position angle 
we assume
the projection of the vector $\vec \Omega$ on the plane of the sky as the
reference direction corresponding to $\psi=0$.
The position angle determined from observations is conventionally
measured from the northern direction through east.
To compare the absolute values of the modelled position angle 
with the data it is therefore necessary to add to the modelled values
the observed position angle of the pulsar rotation axis.
Moreover, we assumed that the polarization direction is \emph{parallel}
to the local acceleration vector $\acc$. If the actual polarization
direction is \emph{perpendicular} to $\acc$ (as in the case of the radio
emission from the Vela pulsar, Lai et al.~2001; Helfand et al.~2001;
Radhakrishnan \& Deshpande 2001) an additional $90^\circ$ shift must be
applied to the modelled position angle curve, before absolute
comparison can be performed.

\section{Results}

\subsection{The two-pole caustic model}

\subsubsection{Lightcurves}

In the two-pole caustic model (Dyks \& Rudak 2003)
pulsations result from rotation of a dipole with
roughly uniform photon emission along the entire length of the
last open magnetic field
lines.
For most dipole inclinations 
two strong caustics form on the trailing side of open field line cones
related to both polar caps.
They can be seen in Fig.~\ref{traj60}, which presents 
projection of magnetic field lines onto the $(\phi, \zobs)$ space,
where $\phi$ is the phase of the pulse, and $\zobs$ is the angle between
the rotation axis and the observer's line of sight.
The pattern was calculated for $\alpha = 60^\circ$, $\prot = 0.033$ s, and
for 90 magnetic field lines with their footprints spaced
uniformly \emph{along the rim} of the polar cap 
as shown in Fig.~\ref{inelpos2} (ie.~non-uniformly in $\phim$).
The blank deformed ovals are polar caps and the
two trailing caustics can be identified as the
dark arches spanning a large range of $\zobs$ near the phases $\phi=0.05$ 
and $0.55$.

The formation of the trailing caustics was explained already by Morini
(1983) (see his fig.~2):
an increase in photon emission altitude along a trailing magnetic field line
results in a phase delay/shift which is almost completely
compensated by effects of aberration and time of flight.
Therefore, photons emitted within a broad range of altitudes
are piled up at roughly the same phase.
On the leading part of the open field line cones
the same special relativity effects produce the opposite effect:
photons are spread out over a large range of $\phi$, forward in phase.

Morini's model assumed that an observer views caustic emission from only one
pole. Dyks \& Rudak (2003) have introduced a model where the observer
sees caustic emission from both poles.
Since for observer positions $\zobs$ departing from the rotational equator
the two trailing caustics approach each other,
observers located at $90^\circ-\alpha \la \zobs \la 90^\circ+\alpha$ 
(and $\zobs \ne 90^\circ$)
will detect double peaked pulse profiles
with the peak separation $\delp \ne 0.5$. 
Observers located at small angles with respect to the rotation axis
($90^\circ+\alpha \la \zobs \la 90^\circ-\alpha$)
will detect single pulse profiles.

As can be seen in Figs.~\ref{traj60} and \ref{rev60}
these conclusions, drawn by Dyks \& Rudak (2003) based on results
for a static shape dipole, prevail in the retarded dipole case.
Fig.~\ref{rev60} presents radiation characteristics predicted by the
two-pole caustic model for a pulsar with $\alpha = 60^\circ$, 
observed at nine different angles $\zobs$.
The other parameters of the model were: $\rovc = 1$, $\rhomax = 0.75 \rlc$, 
$\rmax = \rlc$, and $\prot=0.033$ s.
The figure consists of 9 three-panel frames, 
corresponding to the 9 different viewing angles
 $\zobs$, shown in the upper right corner of upper panels.
Each frame presents     
a lightcurve (upper panel), a position angle
curve (dots, often merging into a thick solid line, middle panel),
and a degree of polarization (thick solid line, lower panel).
In the middle and in the lower panel, the lightcurve from the upper
panel is overplotted for a reference (thick grey line). 

For large viewing angles ($|90^\circ-\zobs| \la \alpha$) 
two peaks (separated by $0.35 - 0.5$) due to crossing the trailing
caustics can be easily identified in the upper panels of Fig.~\ref{rev60}.
The leading peak (P1) is located close to $\phi = 0.1$, ie.~it lags 
by $0.1$ the phase zero, at which an observer's line of sight approaches
most closely the dipole axis.
This is consistent with the relative positions of gamma-ray and 
radio peaks observed for the Vela pulsar and PSR B1951$+$32 (Fierro
1995), since the observed radio peaks are commonly interpreted
in terms of a narrow cone of radiation aligned with the magnetic
dipole axis. The Crab pulsar is a special case with a very 
complicated radio pulse morfology (Moffet \& Hankins 1999).
However, polarization properties of the Crab
suggest that the closest approach to the dipole axis also
occurs at the phase preceding by $\sim 0.1$ the leading gamma-ray peak 
(Moffet \& Hankins 1999), which again agrees
with the prediction of the two-pole caustic model. In section \ref{altitudes}
we show that positions of the main features of the Crab's radio profile 
(a precursor, a main peak, and an interpulse) can be interpreted
within the two-pole caustic model.
In the case of the Geminga pulsar no stable radio profile has been
determined so far (eg.~Malofeev \& Malov 2000) 
and a gamma-ray lighcurve features two peaks
separated by 0.5. The two-pole caustic model pedicts $\delp=0.5$ for
$\alpha\simeq90^\circ$ and any $\zobs$, or for $\zobs\simeq90^\circ$
and any $\alpha$ (Fig.~\ref{rev60}i).
This interpretation contradicts models of Geminga's radio
emission which require nearly aligned
geometry (small $\alpha$ \emph{and} $\zobs$, Gil et al.~1998; Malofeev
1998) but may explain why Geminga's radio flux is so small, since
the line of sight does not approach near to the pole..

Relative intensities and widths of the leading peak (P1) and 
the trailing peak (P2) (Fig.~\ref{rev60}) are far from those
observed in gamma rays (Kanbach 1999, Thompson 2001), however,
we show below that these characteristics are very sensitive
to the magnetic field structure close to the light cylinder,
and to other model parameters.
Apart from the two peaks, there are a few more features 
(marked with the letters O, N, and S in Fig.~\ref{rev60})
present
in the lightcurves. 
Their origin is the following:
Features marked with O arise due to partial overlap of two
emission patterns from each of the polar caps, and can be easily
understood by inspecting Fig~\ref{traj60}.
Features marked with N (Fig.~\ref{rev60}e, f)
are produced when the line of sight crosses
emission from the bunch of lines emerging from the notch part of the
polar cap rim. 
In spite of the fact that the spatial spread of the last open 
magnetic field lines at
high-altitudes is the largest just for the lines
anchored at the notch, projections of these widely spreaded
``notch lines" on the $(\phi, \zobs)$ plane overlap.
This ``notch bunch" of magnetic field lines can be identified
in Fig.~\ref{traj60} in regions which precede the polar caps in phase.  
Had the notch been not present in the polar cap, 
the N-features would disappear.
Features marked with S (small step-like drops in emission level,
Fig.~\ref{rev60}g, h, i)
are produced because high altitude spread of the last open lines
changes discontinuously at the second critical point C2 at the polar cap
rim (even though the lines' footprints are 
spreaded uniformly along the polar cap rim). This slight jump in the lines'
spread can also be noticed in Fig.~\ref{traj60}.\\
Further discussion of lightcurves predicted by the two-pole caustic model
is included in the following subsections.

\subsubsection{Polarization}
\label{caupa}
 
Polarization data on pulsars are limited to radio frequencies, with the
only exception being the Crab pulsar, for which polarization data in 
near-infrared, visible, and ultraviolet light are available
(Jones et al.~1981; Smith et al.~1988; Graham-Smith et al.~1996;
Romani et al.~2001; Kanbach et al.~2003). 
X-ray and gamma-ray experiments will be able to
measure polarization of bright pulsars in the near future (eg.~Bloser
at al.~2003).

The variations of position angle observed for the Crab at optical 
(Fig.~\ref{data}; Kanbach et 
al.~2003) differ from those observed at radio
wavelengths (Moffet \& Hankins 1999).
There may be two reasons for such difference: 
1) Radio emission at fixed frequency is usually
interpreted as coming from a limited range of altitude, 
contrary to the high energy
emission (including optical) which is believed to originate
from a wide range of altitudes,
at least in the two-pole caustic, and in the outer gap model.
Therefore, below we will compare predictions of these models
with the highest energy polarization data which are available, ie. with
the optical data on the Crab pulsar (UV polarization data 
(Graham-Smith et al.~1996) are not as good in quality as the optical
data).
2) Position angle may depend on photon energy.
This is possible because a directional distribution of radiation emitted
by a single electron depends on the electron's energy and, therefore,
on a characteristic frequency of the emitted radiation.
A convolution of these single-electron radiation patterns 
with a spatial distribution of electrons may then result
in the frequency-dependence of position angle (Chen et al.~1996).
This effect certainly is worth closer investigation
but is beyond the scope of the present geometrical approach.\footnote{
Although we largely ignore the spectral dependence of polarization
characteristics, we discern between the coherent
radio emission (with the well established phenomenology)
and the high-energy emission which in general is spatially and
spectrally separated from the radio.
We use the term ``high-energy" for non-thermal X-rays, gamma-rays, and
for optical emission, provided the latter has the same origin as
gamma-rays.
This is justfied in the case of the Crab pulsar, for which both 
optical and gamma-ray emission occur at the same phases and have 
spectra which connect smoothly.}

Perhaps the most striking feature of optical emission from the Crab is
the fixed value of position angle within low-intensity phase
intervals (Fig.~\ref{data}, Kanbach et al.~2003). 
Astonishingly, the value is the same 
for the bridge, and for
the off-pulse region.   
This feature cannot be understood within any
model which connects the position angle directly to the geometry of the
rotating dipole within the light cylinder radius. 
We comment on the possible origin of this constant component in Section
4 (Conclusions).

The constancy of $\psi$ suggests that the optical emission from the
Crab pulsar is a superposition of two components:
1) a pulsed component with highly variable position angle and
polarization
degree, with the latter being very small ($\sim2$ \%)
just behind the peaks' maxima;
2) a constant intensity component with fixed position angle ($\sim 123^\circ$)
and relatively high polarization degree ($\sim 40$ \%).

Such a superposition of two polarized sources 
may easily produce artifact features
in the position angle curve and in the polarization degree curve,
especially when the intensities of the polarized part of the radiation for 
both these
components are comparable.
Unfortunately, the latter condition seems to be the case for the Crab:
although the intensity at the leading peak exceeeds by a factor of $\sim
100$ the intensity of the constant component, the latter seems to be highly
polarized ($\perc \sim 35$ \%), whereas the polarization degree of the pulsed
component drops to $\sim 2$ \% at the leading peak. 
Therefore, even at the maximum of the leading
peak, the polarized intensity of the pulsed component exceeds the
polarized intensity of the constant component only by a factor of 3 --
5.
This makes extracting the pulsed signal from the total data a crucial
task.

The polarization properties of the pulsed
signal are presented by Kellner (2002), who has subtracted 
from the total signal a constant
component with $\psi = 123^\circ$, $\perc=33$
\%, and $I=1.24$ \% of the maximum intensity at the first peak.  
A result of such a manipulation is also shown in the right column of
Fig.~\ref{data} (this paper):
the fast swings of position angle at both peaks
extend over much larger range of $\psi$, and the constancy of $\psi$
beyond the peaks disappears.
The polarization degree assumes low values ($\perc \la 10$ \%)
at all phases dominated by the pulsed signal.
The value of $\perc$ is especially low ($\perc \sim 2$ \%) 
within the phase intervals
trailing both peaks.

Dyks \& Rudak (2003) suggested that the minima in polarization
percentage $\perc$ could naturally result from the caustic nature of the
peaks: superposition of emission from different altitudes
(ie.~with different position angles), which produces the peaks,
could ensure both the decrease in $\perc$ and the fast swings of $\psi$.
This phenomenon was expected for both the two-pole caustic and the outer gap
model.

As can be seen in
Fig.~\ref{rev60} (middle panels, $\zobs \la 70^\circ$)
the two-pole caustic model does predict fast changes of $\psi$
at the leading peak (P1), close to phase $0.1$. 
The change of $\psi$ at P1 is faster on the trailing
side of the peak, than on the leading side, in agreement with the Crab
data.
In the case presented in Fig.~\ref{rev60}, 
ie.~for photon emission limited to the last open magnetic field
lines ($\rovc = 1$), the two-pole caustic model does not predict
the fast increase of $\psi$ at the trailing peak (P2)
located near the phase $\phi=0.5$.
Just behind P2, the position angle changes discontinuously
by a few tens of degrees, when the line of sight starts to sample
emission from the other magnetic pole. 

The fast increase of $\psi$ at the leading peak P1 lags in phase 
the near-pole, radio-like fast swing of position angle at the phase zero
(cf.~Fig.~\ref{rev60}, middle panels). 
The radio-like swing near phase zero has the same origin
as the swings commonly observed at radio frequencies, ie.~it results
from the line of sight passing nearby the magnetic pole. It differs from
the radio swings only in that it is produced by radiation originating
from various altitudes. Like the radio swings, the swing
near the magnetic pole at $\phi=0$
changes sign for the line of sight passing on opposite sides of the magnetic
pole, ie.~it depends on the sign 
of the impact angle $\betmin=\zobs-\alpha$
(compare middle panels of frames d and g in Fig.~\ref{rev60}).
The fast change of $\psi$ which occurs at the leading caustic peak,
however, does not depend on the sign of the impact angle. At P1 the
position angle $\psi$ always increases (for $\zobs < 90^\circ$),
regardless of whether the line of sight passes above, 
or below the magnetic pole.
Thus, the caustic model predicts that the Crab pulsar is viewed at the
angle $\zobs < 90^\circ$. (For $\zobs > 90^\circ$ the PA would decrease
at the leading peak, cf.~Section \ref{conventions}).

In the regions where the caustic pile up is most pronounced
(see Fig.~\ref{traj60}) the geometry of different field lines which
contribute at the same phase is very similar and, therefore, there is no
significant depolarization at the peaks (Fig.~\ref{rev60}, lower
panels). The only considerable drops in $\perc$ occur at the overlap
regions (marked with O in Fig.~\ref{rev60}).
The minima caused by the overlap often occur at (or just behind) the peaks, 
in agreement with observations.

We find that under the present assumptions 
(in particular in the case of the photon emission constrained only to
the last open field lines) the two-pole caustic model
is not able to \emph{exactly} reproduce 
the Crab polarization data, for any model parameters.
Fig.~\ref{effort1} presents results for another type of calculation
-- with electron density distribution slightly smeared
around the polar cap rim (a Gaussian with
$\sigma=0.025$, centered at the rim, between
 $\rovcmin = 0.95$, $\rovcmax = 1.05$).
It has been obtained for $\alpha=70^\circ$, $\zeta=50^\circ$,
$\rhomax = 0.8 \rlc$, $\rmax = \rlc$, and $\prot=0.033$ s.
 
The spread in electron density is responsible for comparable intensity
of both peaks in the lightcurve (Fig.~\ref{effort1}, upper panel).
This is because in the case $\rovc=1$ (Fig.~\ref{rev60})
the leading peak 
is much broader than the trailing peak.
Photon emission from magnetic field line cones with different, fixed $\rovc$
produces peaks at slightly different phases. Broad leading peaks for different
cones overlap in phase, whereas narrow trailing peaks 
do not sum up in phase.
Another new feature caused by the surface density spread
is the fast increase in position angle \emph{behind} the trailing peak
(Fig.~\ref{effort1}, middle panel).
Although resembling the fast swing at P2 observed in the optical Crab data,
the modelled swing does not coincide in phase with P2, and occurs within
the low-intensity trailing wing of P2.
Moreover, 
at the leading peak the modelled position angle $\psi$ spans
a different range of values ($50^\circ - 180^\circ$) than at the trailing wing 
of P2 ($-50^\circ - 90^\circ$). The observed range of values assumed by the
position angle at the leading peak of the Crab's optical profile
is roughly the same as the one observed at the trailing peak
(Fig.~\ref{data}).
The last effect of the density spread is
the noticeable decrease in the degree of polarization near the peaks
(Fig.~\ref{effort1}, bottom panel).
The narrow dip in polarization degree which almost coincides with P1
(located near $\phi=0.1$) is caused by superposition of emission
from nearby magnetic field lines (with slightly different $\rovc$) 
emerging
from the same, northern magnetic pole.
Likewise, the minimum in $\perc$ at the phase $\phi=0.48$ 
which slightly precedes the trailing peak originates from 
the superposition of emission from nearby magnetic field lines,
all of which
emerge from the same, southern magnetic pole.
Although the superposition of emission from adjacent magnetic field lines
occurs at any phase $\phi$, 
it results in the drop in $\perc$ only at the peaks.
This indicates that
the caustic effects at the peaks are crucial for producing this minima in
$\perc$.
The other minima in $\perc$, following both peaks 
result from superposition of radiation patterns from the different 
magnetic hemispheres.
Their shape and depth depends sensitively on the parameters
$\rhomax$ and $\rmax$, since including the near-$\rlc$ region greatly
increases the area on the $(\phi, \zobs)$ plane within which
the two radiation patterns from the opposite magnetic hemispheres 
overlap. Therefore, the degree of polarization may provide a good
diagnostic of pulse profiles with overlapping emission
from regions widely separated in space.
There is a large 
variety of polarization degree curves which can be produced
by the two-pole caustic model, given the large number of 
combinations of parametrs $\alpha$, $\zobs$, $\rhomax$, $\rmax$.
We find that in general, however, both the leading and the trailing peak
are followed by a minimum in polarization degree, which
is in qualitative agreement 
with the optical data on the Crab pulsar (Fig.~\ref{data}, bottom right
panel).

\subsubsection{Static shape dipole}

A rigidly rotating static shape dipole is worth consideration for two
reasons: 1) calculations are much simpler for this field structure;
2) by comparison of results for the static dipole with those for the
retarded dipole one can assess the importance of the near-$\rlc$
distortions of magnetic field for predicted radiation characteristics.

\begin{figure}
\epsscale{0.7}
\plotone{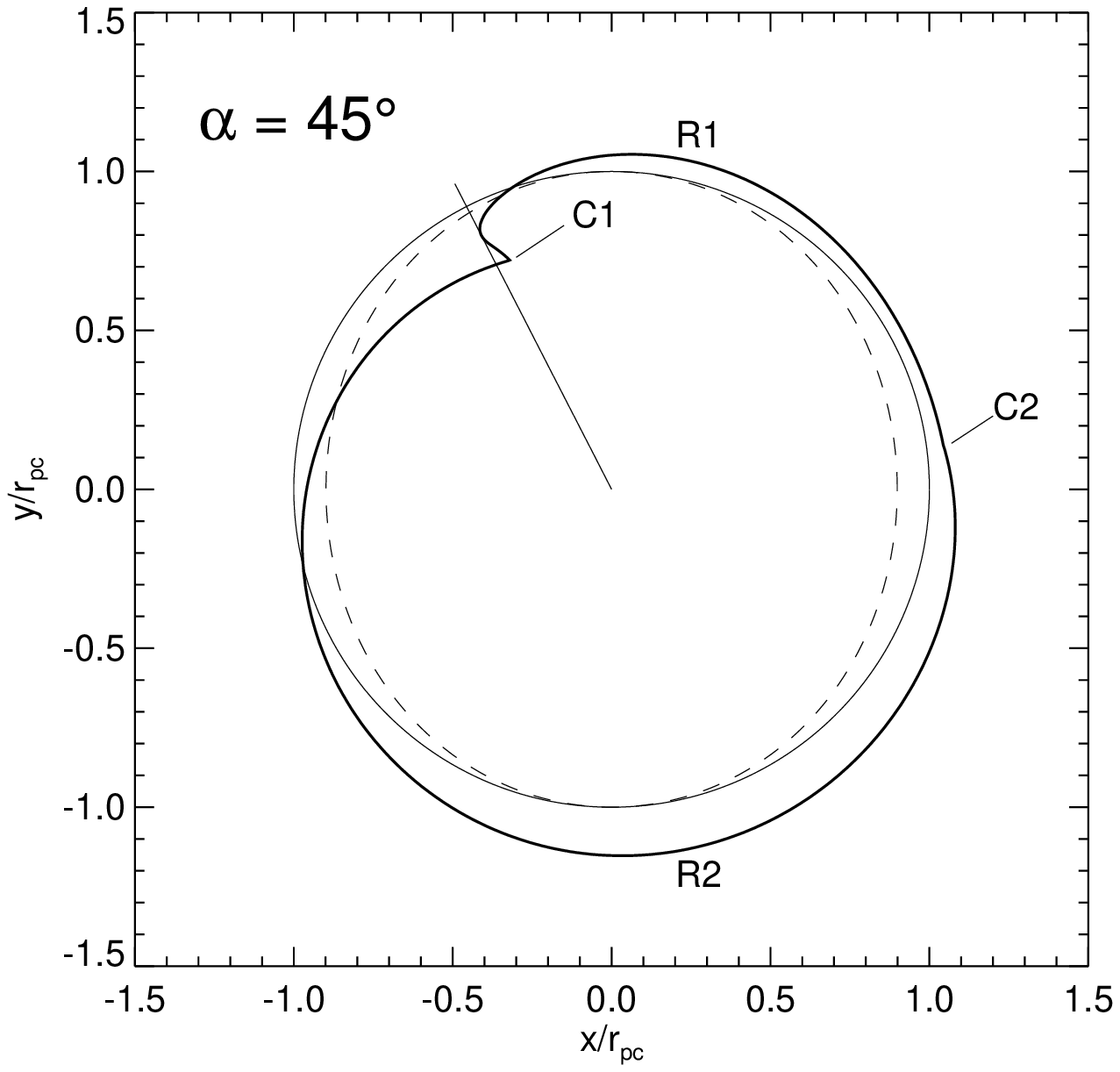}
\caption{Two polar caps for a dipole tilted at $\alpha = 45^\circ$
and rotation period $\prot=0.033$ s: 
Thick solid line (with a notch) presents the polar cap rim 
for the retarded dipole. 
The dashed oval is for the static shape dipole.
A circle of radius $\rpc$
(corresponding to a polar cap of a dipole aligned with rotation axis)
is added for a reference (thin solid).
Note that for most azimuths,
the circle approximates
the retarded case (thick solid) better than the actual rim of the cap 
for static dipole (dashed).
The magnetic azimuth $\phim$ is measured counterclockwise from positive
$x$ axis. The closest rotational pole is on the left ($\phim =
180^\circ$). The corotational velocity points upwards ($\phim =
90^\circ$). In this figure, as well as in Figs.~\ref{rings2} and 
\ref{inelpos2}, the magnetic moment $\vec \mu$ of the retarded dipole 
is aligned with the magnetic moment of the static dipole 
($\vec \mu$ is located  perpendicularly to the page at the point $(x,y)=(0,0)$).
\label{caps2}}
\end{figure}

Let us first discuss the static shape dipole with photon emission
from ``actual" last open magnetic field lines, ie.~those which are 
tangent to the light cylinder and emerge from the oval-shape polar cap
rim, drawn with the dashed line in Fig.~\ref{caps2}.
We find pronounced differences between radiation characteristics calculated
for this case and for the above-described retarded case.
In the static dipole case, and for moderate inclination
angles ($\alpha\sim 60^\circ$), the radiation pattern from a
single magnetic pole
tends to be projected mostly within one rotational
hemisphere -- ie.~when projected on the $(\phi, \zobs)$ plane
of Fig.~\ref{traj60} it would mostly occupy only its upper half 
(or lower half, for
the opposite pole). 
There is a very small range of viewing angles close to the rotational
equator ($\zobs \ga 80^\circ$), for which the pulse profiles are
double-peaked and resemble those shown in Fig.~\ref{rev60}.
It is difficult to obtain peak separation 
much different from $0.5$ and
most viewing angles result in single peak lightcurves.
These modifications take place because the last open magnetic field lines 
of the static shape dipole
emerge from the oval-shape polar cap rim -- squeezed in the direction of
rotational colatitude $\theta$ 
(Fig.~\ref{caps2}, dashed line). Magnetic field lines anchored at 
the oval rim close to $\phim
=0$ or $180^\circ$ extend in the $\theta$ direction much less than
than those anchored at the same $\phim$ in the retarded dipole case
(cf.~the shapes of polar cap rim for these two cases, Fig.~\ref{caps2}).

We conclude that the rotational distortions of the near-$\rlc$ magnetic field 
structure have a large influence on the resulting radiation
characteristics.
We emphasize that
the distortions of the pulsar radiation pattern
are effects of \emph{lower} order than $\betcor^2$, where $\betcor$ is a local
corotation velocity in units of the speed of light. Eg.~the average radius
of the polar cap shown in Fig.~\ref{caps2}
changes by $\sim 10$ \% (or, by $\sim 8\cdot10^{-3}$ rad), 
whereas $\betcor^2=2\cdot10^{-5}$ for $\prot=0.033$ s,
$\rns=10^6$ cm, and a
distance of the polar cap from the rotational axis 
$\rns\sin\alpha\simeq 7\cdot 10^5$ cm.
In fact, the rotational distortion of the radiation pattern is even 
larger than $\betcor$ which is equal to $4.5\cdot10^{-3}$ at the polar cap
shown in Fig.~\ref{caps2}.
Some geometrical models of radio properties of pulsars (eg.~Blaskiewicz
et al.~1991; Gangadhara \& Gupta 2001) fully rely on symmetry of
the open field line cone with respect to the $(\vec \Omega, \vec \mu)$ plane.
The rotational distortions of the magnetic field
are neglected in the models, and
the asymmetry of radio pulse profiles with respect to some
fiducial features (eg.~the center of position angle curve, Blaskiewicz
et al.~1991, or the position of the core component, Gangadhara \& Gupta 2001)
is interpreted purely in terms
of aberration and finite propagation time of radio waves.
The strong, rotational distortions of the open field line region, 
visible in Fig.~\ref{caps2}, indicate that the rotationally induced
sweep-back of magnetic field lines should have been taken into
account in these models.

The most convenient calculation of pulsar radiation characteristics
is based on the static shape dipole geometry, \emph{and} it employs the
standard circular polar cap rim of radius $\rpc$ as the position of
footprints of the ``last open" magnetic field lines at the star surface.
Contrary to the above discussed case of the oval cap, 
a calculation of this type approximates very well the
lightcurves obtained for the retarded dipole, a fact emphasized
by RY95.
This can be easily understood:
as can be seen in Fig.~\ref{caps2} the circular polar cap
approximates the retarded polar cap 
better than the actual, oval-shape polar cap for 
the static dipole case (cf.~thin solid, thick solid, and dashed polar
cap rims in Fig.~\ref{caps2}).

Fig.~\ref{effort2} presents a lightcurve (a), a position angle curve
(b), and
a degree of polarization (c) 
for the static dipole case with the circular polar cap.
Parameters used in the calculation were the same as in
Fig.~\ref{effort1}.
Note the resemblance of the lightcurve 
and the polarization characteristics to those one
for the retarded dipole case (Fig.~\ref{effort1}).

Reproducing the lightcurve of the Crab pulsar is a little bit
problematic for the two-pole caustic model, because it often predicts
a bump of emission at phases following the first peak
(Fig.~\ref{rev60}c-g, Fig.~\ref{effort1}a).
This feature has a twofold origin:
1) in part it is produced by an overlap of emission patterns
from two poles;  
2) it is a trailing wing of a peak which would form very close to the
light cylinder ($\rho > 0.8 \rlc$) at the leading side of the open field
line cone. This is the peak which is interpreted as the leading peak
in the outer gap model (RY95; CRZ2000). It usually slightly lags the
first peak of the two-pole caustic model, although the actual value of the
phase lag depends on viewing geometry. 
In the lightcurves predicted by the two-pole caustic model 
the peak appears only 
when high-altitude emission is included. Had the photon emissivity
declined with altitude above some $r \ga 0.5 \rlc$, the feature would be much
less pronounced or would disappear.

Interestingly, however, such a feature is present in the gamma ray lightcurve
of the Vela pulsar (Grenier et al.~1988; Kanbach et al.~1994; Thompson 2001).
Fig.~\ref{effort3} shows the Vela lightcurves for photon energy range
of $2.8-10$ GeV (a) and for the entire energy band of EGRET
($30$ MeV -- $10$ GeV, panel b).
A lightcurve predicted by the two-pole caustic model is shown in panel c. 
Note that the shape of the broad-band lightcurve is very well
reproduced by the two-pole caustic model: 
the leading peak is narrower than the
trailing peak, which connects smoothly with the bridge emission.
The leading peak does not connect smoothly with the bridge, and is
followed by the ``postcursor bump".
These features result naturally from the two-pole caustic model, since it
predicts that the trailing peak, the bridge emission, and the
``postcursor bump" arise from sampling a single, continuous radiation
pattern from one magnetic pole
(cf.~Fig.~\ref{traj60}). The leading peak and the offpulse
emission are produced by sampling an emission pattern from 
the opposite magnetic pole.
Such a two-component decomposition of the Vela lightcurve is confirmed by the
lightcurve observed within higher energy range between $2.8$ and $10$ GeV
(Fig.~\ref{effort3}a), where the trailing peak remains
smoothly connected to the bridge, whereas the leading peak seems to
present a separate entity.  
We emphasize that our interpretation of the gamma-ray lightcurve of the
Vela pulsar (with $\alpha\simeq 70^\circ$ and $\zobs\simeq 60-65^\circ$) 
is in very good agreement with a geometrical
model of X-ray arcs surrounding the pulsar in Chandra images
(Radhakrishnan \& Deshpande 2001 -- the model; Pavlov et al.~2000 and
Helfand et al.~2001 -- observations). In Fig.~4 of Dyks \& Rudak (2003) 
we chose $\alpha = 70^\circ$ and $\zobs = 61^\circ$ 
to reproduce the observed peak separation of $\sim 0.43$ while keeping the
difference $\alpha - \zobs$ smaller than $\sim 10^\circ$ to ensure
the close approach to the narrow radio beam centered at the dipole axis. 
Given the relatively
weak sensitivity of the peak locations to $\alpha$ and $\zobs$
(eg.~cf.~Fig 3 in RY95) the inferred values of these parameters
have accuracy of a few degrees. 
Radhakrishnan \& Deshpande (2001) infer similar values of
$\alpha=71^\circ$ and $\zobs=65^\circ$ 
from their interpretation of the X-ray arcs, ie.~on a
completely different basis than our. 
We have used their values to calculate the two-pole caustic model
lightcurve in Fig.~\ref{effort3}c.
The closer approach to the dipole axis
($|\zobs - \alpha| = 6^\circ$ in comparison with $9^\circ$ in Dyks \&
Rudak 2003) is in excellent agreement with the fits of the position angle
swing to the radio data (Krishnamohan \& Downs 1983;
Johnston et al.~2001).

The model results in Figs.~\ref{effort2} and \ref{effort3}c 
have been calculated for the
circular-cap approximation and for the static-shape dipole, which are
also assumed in Section 3.3 and in Fig.~\ref{polcap}.
All the other results presented in this paper have been obtained
for the retarded dipole with the distorted shape of the polar cap.

\subsection{Outer gap model}

In this section we decribe predictions of the most popular version
of the outer gap (OG) model described eg.~in RY95 and CRZ2000.
This version assumes that the outer gap extends from the null charge
surface up to the light cylinder not only in the $(\vec \Omega, \vec
\mu)$ plane, and that radiation emitted outward
dominates. 
Other versions of the OG model (Cheng et al.~1986; Chen et al.~1996)
are not treated here.

Although the traditional geometry of the outer gap 
has recently been questioned
(Hirotani \& Shibata 2001) the model maintains its great popularity
because it naturally explains some features of observed
lightcurves (in particular the relative positions of gamma-ray and radio
peaks, Chiang \& Romani 1992; RY95)
and it \emph{seemed} to produce fast swings of position angle at both peaks
(RY95).

Our calculations confirm the success of the OG model
in reproducing the relative phases of radio and gamma ray peaks;
however, 
our polarization results for this model 
do not agree with the data and
contradict previous results presented by RY95. 
Moreover, the ability of the model to reproduce lightcurves
fully relies on including photon emission from the very vicinity
of the light cylinder ($0.8\rlc \la \rho \la \rlc$), 
where the vacuum magnetic field dipole
approximation is not valid.

In the outer gap model, the position of magnetic field line footprints
at the star surface becomes a crucial parameter.
For the simplest choice: $\rovc = 1$ (polar cap rim) the observed
gamma-ray lightcurves cannot be well reproduced and, therefore, 
it is necessary to
assume $\rovc < 1$ (usually close to $0.9$).

Fig.~\ref{revog} presents lightcurves and polarization calculated
for the outer gap model with $\alpha = 65^\circ$,
$\rhomax = 0.999\rlc$, and $\rmax = 1.7\rlc$.
A gaussian spread in electron density has been assumed 
in this calculation with $\rovcz=0.9$, 
$\sigma=0.025$ and $\rovc$ ranging between $0.85$ and $0.95$.
The results are presented in the same way as in Fig.~\ref{rev60}.
For large viewing angles $\zobs$ (Fig.~\ref{revog}g, h, i)
the well known double peak shape of lightcurves 
obtained by RY95 and CRZ2000 can be recognized.

A relatively fast increase of the position angle appears at the trailing peak
(Fig.~\ref{revog}e, f, g, h). At the leading peak, however, either
the fast changes of PA do not occur (Fig.~\ref{revog}g, h, i)
or they occur in the opposite direction than at the trailing peak
(Fig.~\ref{revog}e, f).
The same behaviour also takes place for larger dipole inclinations $\alpha$,
including $\alpha \ga 70^\circ$ for which RY95 found fast changes of
$\psi$ at both peaks (for $\zobs \la 70^\circ$).

Our polarization results for the outer gap model
cannot reproduce those published by RY95. We have checked if the disagreement
results from different calculation methods -- RY95 assumed
that the polarization direction $\vec E_{\rm w}$ is parallel to the local
radius of curvature of magnetic field lines $\vec \rho_{\rm curv}$, 
whereas we present the results for $\vec E_{\rm w}\parallel \vec a$.
We find that for $\vec E_{\rm w}\parallel \vec \rho_{\rm curv}$
the fast changes of PA do occur in the parameter range suggested
by RY95 ($\alpha \ga 70^\circ$, $\zobs \la 70^\circ$), however, only
at the leading peak. Moreover, we find that even the swing at
the leading peak is problematic, since it occurs for photon
emission along the magnetic field lines emerging very close to the polar
cap rim ($\rovc \ga 0.9$) whereas the good looking lightcurves
are produced for ($\rovc \la 0.9$).
Although we used the same calculation method, the same model parameters
as in RY95 and we densely sampled a large range of model parameters
not specified in RY95, 
we were not able to reproduce the result shown in fig.~5 
of RY95.\footnote{Unfortunately, RY95 
do not specify what
value of $w$ was used in their calculation. 
We were not able to reproduce their result for any value of $w$. 
We argue that either their result is incorrect, or it requires
a very precise adjustment of model parameters. Certainly, 
it is not a natural prediction of the outer gap model.}

A serious disadvantage of the outer gap model is that the swing at
the leading peak (as well as the peak itself -- see below) is formed
very close to the light cylinder -- at $0.8\rlc \la \rho \la \rlc$ (and
$r \ga \rlc$), where the geometry of the magnetic field is not known. 
In calculations which discard the near-$\rlc$ emission ($\rhomax = 0.8\rlc$), 
the leading peak is missing.

\subsection{Polar cap model}
\label{polar}

In the polar cap model (Ruderman \& Sutherland 1975; Arons \&
Scharlemann 1979)
acceleration takes place close to the star surface
(at altitudes $h \la \rpc$ or $h \la \rns$, depending on the boundary
conditions at the surface), which implies a very narrow beam for
gamma-ray, and radio emission.
Therefore, the polar cap model
is able to reproduce the widely separated double peaks in pulsar
lightcurves only if the rotation axis, the magnetic dipole
axis, and the observer's line of sight are nearly aligned
(Lyne \& Manchester 1988; Daugherty \& Harding 1994; Dyks \& Rudak
2000).
With the nearly aligned geometry included, the model is able to
reproduce phase-resolved high-energy spectra of the Vela pulsar
(Daugherty \& Harding 1996) as well as a fading of the leading peak at
the high-energy spectral cutoff (Dyks \& Rudak 2002).
A long list of advantages of the polar cap model over the other models
can be found in Baring (2001).

Slot gap geometry somewhat eases the requirement of alignment, but small
inclination angles are still necessary (Muslimov \& Harding 2003a).
This is a real
problem for the polar cap model, since there is no 
statistical evidence for more frequent occurence of small inclinations 
angles among young pulsars
(Lyne \& Manchester 1988; Blaskiewicz et al.~1991).
The polar cap model predicts that the nearly aligned rotators are about
one order of magnitude brighter gamma-ray sources 
than pulsars with larger dipole
inclinations (Dyks 2002). This is because in the nearly aligned
geometry, the observer's line of sight samples the high intensity polar cap
beam of radiation for a much longer fraction of the rotation period 
than in the case of
large dipole inclinations. Although this implies higher probability 
of detection for the nearly aligned rotators, X-ray and radio studies of
particular cases 
contradict the nearly aligned
geometry
(eg.~Crab, Hester et al.~1995, and Vela, Krishnamohan \& Downs 1983; 
Radhakrishnan \& Deshpande 2001). 
Moreover, although the near alignment ensures large separations between
the two gamma-ray peaks, it does not explain why so many gamma-ray
pulsars ($\sim 50$ \%) have $\delp$ in the narrow range 
between $0.4$ and $0.5$.
Apparently, the polar cap model is not able to explain the observed
shape of gamma-ray lightcurves naturally.

Moreover, the polar cap model predicts that the only fast swing in
position angle $\psi$ should occur midway between the gamma-ray peaks
-- not at the phase of the gamma-ray peaks.
A typical shape of the position angle curve predicted by the polar cap model
is shown in Fig.~\ref{polcap}b. 
It has been calculated for the static shape dipole geometry of the
magnetic field with the circular polar cap.
The following, simplified emission region was assumed to calculate
the PA curve:
For $\rovc < 1$ (ie.~inner regions of the open field line tube)
the high-energy radiation was assumed to be emitted along all 
(open) field lines at a fixed radial distance $r=3\rns$.
For $\rovc = 1$ (the last open field lines) the emission originated
from all radial distances exceeding $3\rns$. 
This choice of the emission region is an effort to simplify the
complex emission pattern of the polar cap model:
At any rotational phase, an observer detects radiation from many
different altitudes and from many different magnetic field lines. 
However, as long as the line of sight cuts through the acceleration
region 
located at $r=3\rns$ (which takes place for $|\phi|\la 0.2$ for the viewing
parameters used in Fig.~\ref{polcap}) the high-energy
radiation from the accelerator should dominate the detected signal.
For $|\phi|\ga 0.2$ the line of sight no longer cuts through 
the accelerator (limited to the open field lines), and, therefore,
the received radiation should be dominated by emission from the last
open magnetic field lines,
 where the model assumes an enhanced electron density to produce pronounced 
peaks.

The central parts of the PA curve, due to the photon emission
from the fixed altitude, are marked in Fig.~\ref{polcap} with dots.
The fast swing of $\psi$ visible in Fig.~\ref{polcap}
near $\phi = 0$ has the same origin as the
swings observed at radio frequencies, ie.~it results from the line of sight 
passing nearby the magnetic pole.
Contrary to the prediction of the polar cap model, 
the well established interpretation of position angle
swings at radio frequencies implies that the swing should occur at 
the phase of a radio peak.
In the case of pulsars with gamma-ray profiles with two
widely separated peaks (Crab, Vela, B1951$+$32) the phase of the radio
swing precedes by $\sim 0.1$ the leading gamma-ray peak.
If the standard interpretation of radio properties applies to gamma-ray
pulsars (and, except from the Crab case, there is no reason for which
it should not apply), it contradicts the 
polar cap model unless the radio peaks are leading edge cone emission.

As described in BCW91 and HA2001, the special relativity effects
associated with the star's rotation influence the position of the PA
curve. Two effects of rotation can be discerned in Fig.~\ref{polcap}:
1) the delay of the PA center (located at $\phi\simeq 2r/\rlc$) 
with respect to the profile center (at $-2r/\rlc$) discovered by BCW91, 
and 2) the upward shift of the PA curve by $(10/3)(r/\rlc)\cos\alpha$
found by HA2001 
(the \emph{downward} shift, as originally described by HA2001, takes place for 
$\zobs > 90^\circ$, see the discussion in Section \ref{conventions}).
The presence of these effects in the modelled PA curve confirms
the correctness of our numerical code.
Since in our geometrical approach a depolarization can result only
from overlaps of emission from different altitudes, for the polar cap
model it remains
constant at the ``intrinsic" level of $80$ \% for any phase $\psi$
(not shown in Fig.~\ref{polcap}).  

As long as the low altitude high-energy emission and the nearly aligned
geometry is considered, the
caustic effects are not important for the polar cap model. However,
the effects become important if lower-energy emission (eg.~optical, X-ray)
from high-altitudes and larger dipole inclinations are considered
(an example of such a case is the polar cap model of Morini 1983).
The original polar cap model 
where acceleration is limited by
electromagnetic cascades 
at low altitude induced mostly by curvature radiation, predicts lightcurves 
very similar to those for the
two-pole caustic model (Fig.~\ref{rev60}) in two cases: 
1) in the off-beam viewing geometry;
2) for any viewing geometry 
at photon energies smaller than $\sim 100$ MeV
and only 
within photon energy range dominated 
by curvature radiation (not by the synchrotron radiation).
In addition to the two caustic peaks, the
lightcurves often possess an additional peak
near phase zero, caused by enhanced photon emission near the star
surface.

These lightcurves, however, could be observed only at photon energies
lower than $\sim 100$ MeV, which is the energy of the ``cooling break" 
in the curvature spectrum
(the position of the break in the curvature spectrum does not depend on pulsar
parameters, see Rudak \& Dyks (1999) for details). 
To explain the lightcurves observed by EGRET, however, 
the polar cap model has
to employ the problematic nearly aligned geometry
or it has to abandon the spatial boundaries of the
traditional polar gap, eg.~by considering the slot gap model where
acceleration continues to high altitude along the last open field lines 
(Arons 1983;
Muslimov \& Harding 2003a). Geometrically, however, this makes the polar
cap model similar to the two-pole caustic model. 

We conclude that the standard, low-altitude version of the polar cap model
is not able to naturally explain the observed geometric properties of known
gamma-ray pulsars. 
Given the firm position of the polar gap among the possible acceleration
sites in the pulsar magnetosphere, however, it is still likely that the
polar cap activity does take place, and that the narrow gamma-ray beam 
of the polar cap model
will eventually be detected
by GLAST.

\subsection{Relative positions of radio and gamma-ray peaks}
\label{altitudes}

RY95 showed that the outer gap model is very successful in reproducing
relative positions of
radio and gamma-ray peaks in pulsar lightcurves (RY95).
The two-pole caustic model works equally well for pulsars with wide peak
separations (like Crab, Vela, and B1951$+$32).
Moreover, the models naturally explain why such a large fraction 
of known gamma ray
pulsars ($\sim 50$ \%) exhibit double peak lightcurves with
a very wide separation between the peaks $\delp \sim 0.4 - 0.5$. 

Both these models imply that a single radio peak, which often
slightly precedes in phase the first gamma-ray peak, is either
a low altitude core component 
or (more probably) a cone component 
arising when our line of sight cuts through the edge of a cone
of radio emission
(hereafter 
we assume the terminology used for
the ``multiconal" model of radio emission pattern
(Rankin 1993; Gil \& Krawczyk 1996; 
Gangadhara \& Gupta 2001; Kijak \& Gil 2002), although a radio beam of
\emph{any} shape (eg.~the ``patchy" beam, Lyne \& Manchester 1988)
can be accommodated by the two-pole caustic and the OG model).
The radio pulses of many young pulsars are now thought to be edges of
cones or partial cones (Crawford et al.~2001; Crawford \& Keim 2003).

The interpretation of the radio peaks
as cones is much
more probable than the core interpretation for two reasons:
1) all models of particular cases (Vela, Crab) employ relatively large
impact angles $\betmin = \zobs - \alpha$ (the minimal angles between 
the line of sight and the dipole axis); 
(eg.~RY95 assume $\betmin = -18^\circ$ for the Crab (fig.~5 in RY95) 
and $\betmin \approx 13^\circ$ for the Vela (fig.~2 in RY95);
CRZ2000 assume $\betmin \approx 17^\circ$ (fig.~7 in CRZ2000);
we assume $\betmin \approx 20^\circ$ in Fig.~\ref{effort1} and
$\betmin \approx 6^\circ$ in Fig.~\ref{effort3}, this paper.)
2) in the case of the Crab pulsar the maximum slope of the position angle
curve fitted to the radio data occurs just ahead of 
the low frequency component 
(hereafter LFC, Moffett \& Hankins 1999), which precedes in phase 
the precursor component. The LFC appears to be closer to
the dipole axis than the precursor (Fig.~\ref{radiocrab}), 
which indicates that the
latter is a cone (below we argue that it must be an \emph{inner} cone). 
Unfortunately, this result is uncertain due
to the poor quality of the fit within the main peak.

The Crab pulsar is an especially interesting case, because 
two of its radio peaks (the main peak MP, and the interpulse IP,
Fig.~\ref{radiocrab})
nearly coincide in phase with gamma-ray peaks P1 and P2.
Assuming that this coincidence implies the same emission regions
for radio and gamma-rays, it becomes possible to
determine the location of the radio-emitting regions
within the models of high-energy emission.
Absolute timing analysis of the Crab pulsar performed
recently in X-rays (Rots et al.~2000; Tennant et al.~2001; Kuiper et
al.~2003) as well as in gamma-rays (Kuiper et al.~2001) revealed
that the main radio peak
of the Crab slightly lags in phase (by about $300$ $\mu$s,
ie.~about 1\% of the total phase)
the maximum of the leading high-energy peak. 
This complicates the analysis because in general this lag can be
interpreted in terms of either a different set of 
radio-emitting magnetic field
lines (different $\rovc$) or of different altitudes for the
radio emission. To avoid these complications, we will first neglect the
lag and assume
that the main radio peaks of Crab (MP and IP) coincide with
the two high-energy peaks.
Possible interpretations
of the lag will be given later in this section.

The radiation observed in the radio peaks that are (nearly)
coincident in phase with the high-energy peaks
should originate from the last open magnetic field lines (two-pole
caustic model)
or from lines lying very close to the last open 
field lines (with $\rovc = 0.8 - 1.0$, 
outer gap model). 
Thus, the MP and the IP present the outermost cone, connected
to the polar cap rim.
We emphasize, that both the two-pole caustic and the outer gap model
can accommodate the coincidence of radio and gamma peaks
in two situations:
1) when the roughly uniform radio emission (per unit length of 
a magnetic field line) occurs within the entire acceleration region
of a given model, and 2) when the radio emission regions are constrained
only to those regions of the accelerator where the caustic peaks are
formed.
In the first case, the radio-emission 
would extend along 
the entire length of all last open magnetic field lines (two-pole 
caustic model) or along ``nearly last" open field lines at all altitudes above
the null charge surface (outer gap model).
However, this possibility would require coherent processes 
at extremely different altitudes, and the
radio emission would appear within the bridge and the offpulse
region (the latter in the two-pole caustic model only), 
which is not observed in the radio lightcurves.
In the second case, different radio frequencies can still be radiated
at different altitudes within the caustic region, however, the
radius-to-frequency mapping would not be apparent.

For the trailing peak (P2, or for IP in radio)
the caustic regions are roughly the same 
in both models (two-pole caustic and OG), since the same caustic is responsible
for producing P2 in the models.
We find that in the case of the viewing geometry suited to reproduce
the Crab's lightcurve, the trailing peak (and so the radio IP of the
Crab)
is produced at radial distances in the range $0.4 - 0.8\rlc$ (two-pole
caustic model)
and $0.2 - 0.6\rlc$ (OG model). 
Typical emission altitudes for the models are presented 
in Figs.~\ref{radii1} (two-pole caustic) and \ref{radii2} (OG).

The origin of the main radio peak in the two models is completely
different, as is the emission region for the leading gamma-ray peak.
In the two-pole caustic model the P1 (or MP in radio) 
has exactly the same origin
as P2 (or IP). Therefore, emission altitudes for P1 are only 
slightly lower than for P2 and typically fall in the range $0.3 -
0.6\rlc$ (Fig.~\ref{radii1}). This slight difference results from
a closer approach to a magnetic pole at P1 than at P2
(cf.~Fig.~\ref{traj60}).
In the outer gap model, the main radio peak (and P1 in
gamma-rays)
is due to emission from the very vicinity of the light cylinder
(at $\rho\sim \rlc$, Fig.~\ref{radii2}, and often even at $r > \rlc$).
This is in clear disagreement with estimates of radio emission
altitudes (eg.~Blaskiewicz et al.~1991; Kijak 2001; Gupta \& Gangadhara
2003), although it must be remembered that the Crab does not fit 
in the radius-to-frequency mapping scheme. 
The radio precursor in the OG model comes from the opposite pole
(and from much lower altitudes)
than the main radio peak.

The radio emission only from the caustic regions is more acceptable
in the case of the two-pole caustic model.
The model is able to produce the gamma-ray-coincident radio peaks
for a radio emission region which spans a \emph{single} range of altitudes 
between $0.3 - 0.6\rlc$. If the radio emissivity in this region
had been uniform in $\phim$, a low-intensity
bridge radio emission would be produced in this
model (this may be easily inferred from Fig.~\ref{radii1} by determining
the range of phases with emission altitudes in the range $0.3 - 0.6\rlc$).
To avoid the bridge in radio, one would have to constrain the radio
emission to the trailing part of the open field line cones
(ie.~to the regions which produce the two gamma-ray peaks in the
two-pole caustic
model). This is not unnatural, because radio lightcurves of many pulsars
suggest azimuthal asymmetries in radio emission cones. 
These often include the
lack of a leading part of the emission cone: Lyne \& Manchester (1988)
find that $30$ \% of asymmetric radio profiles consists of the trailing part
of emission cones only. 

In the case of the outer gap model, to avoid assuming 
radio emission from all altitudes above the null charge surface,
\emph{two} regions of radio emission 
are required: 1) between $0.8$ and $1.0\rlc$ (for P1 and MP), and
2) somewhere between $0.2$ and $0.6\rlc$ (for IP and P2).
Had the radio emission from both these regions been uniform in
$\phim$, both of them would produce radio features within the bridge
region (Fig.~\ref{radii2}). To avoid this, one would have to assume that
on the leading side of the open field line cone the radio emission
only occurs close to the light cylinder (region 1), 
whereas on the trailing part only at low altitudes (region 2).   

Thus, to explain the radio MP and IP peaks of the Crab 
within the two-pole caustic model,
it is sufficient to assume that the intensity of radio emission 
is nonuniform in
magnetic azimuth $\phim$, which is not a stringent requirement.
In the case of the outer gap model, it is necessary to assume
that both radio emission altitudes, as well as the intensity of radio
emission are nonuniform in $\phim$. Moreover, in the OG model,
the main radio peak
would have to be produced close to $\rlc$ contrary to the
precursor, which is assumed to originate from low altitudes and from the
opposite magnetic pole than the main peak. 
We conclude that the gamma-ray-coincident radio peaks 
of the Crab pulsar speak in favour of
the two-pole caustic model. 
They can be understood as emission from the last open
magnetic field lines on the trailing part of the ``polar cone",
within the range of altitudes between $\sim 0.3$ and $\sim 0.6\rlc$.
The Crab's precursor and the LFC are interpreted in this picture as
inner conal components, with the LFC probably being the closest to the
magnetic dipole axis.

To interpret the lag of the MP with respect to the high-energy 
leading peak
(Kuiper et al.~2003 and references therein) 
it is necessary to determine $\rovc$ for
the radio-emitting magnetic field lines, which would require detailed
calculations of pair-formation fronts in the considered models.
Making the simplest assumption, however, that the same set of field
lines is responsible for both the high-energy and the radio emission,
one can interpret the lag with the help 
of Figs.~\ref{radii1}c (two-pole caustic model) 
and \ref{radii2}c (outer gap model).
As can be inferred from Fig.~\ref{radii1}c, 
in the case of the two-pole caustic model
the maximum of the high-energy leading peak
corresponds to $r \approx 0.5 \rlc$ and the detection phase
\emph{increases} with increasing altitude 
(for the viewing geometry used in Fig.~\ref{radii1}; 
for $\alpha = \zobs
= 90^\circ$ the detection phase at the trailing peak may \emph{decrease} with
altitude, depending on the considered altitude range, see.~fig.~2
in Morini 1983).
Therefore, 
the maximum of radio emission 
should occur at a slightly \emph{larger} radial distance
than the one corresponding to the \emph{apparent} maximum of high-energy
emission at the leading peak.
Unlike in the case of the two-pole caustic model, 
according to the OG model, near the leading peak 
the phase of detection \emph{decreases} with emission altitude
(Fig.~\ref{radii2}c, lower curve;
the upper curve in Fig.~\ref{radii2}c
should not be considered since it
corresponds to the emission from the very vicinity of the light cylinder). 
As noted by Kuiper et al.~2003,
this implies that the maximum of radio emission
must take place at a slightly \emph{smaller} radial distance than the one which
corresponds to the maximum of the observed high-energy leading peak
($r \approx 0.9 \rlc$).

We emphasize that what we have just determined from 
the radio-to-high-energy lag is the \emph{absolute} location
of the radio emission region 
(which is $r\approx 0.5\rlc$ at the trailing side of the open field line 
region
for the two-pole caustic model, 
and $r\approx 0.9\rlc$ at the leading side of the
open volume for the OG model)
--- not the relative locations of the strongest
high-energy and the radio emission regions. 
One should not, therefore, conclude that in the OG model 
the strongest X-ray emission takes place
at larger altitudes than the strongest radio emission 
because in both models (two-pole caustic and OG) the high-energy
emission is assumed to be uniform per unit length of a magnetic
field line within a large range of altitudes, and the observed
high-energy peaks are of purely caustic origin.
In the case of both models, the spatially constrained region of 
the strongest radio emission
is contained entirely within the extended region of uniform
high-energy emission.

Pulsars with two widely separated gamma-ray-peaks but without gamma-ray
coincident radio peaks (like the Vela or B1951+32)
apparently do not have the outermost radio cone related to the last
open lines. The single radio peak preceding the leading gamma-ray peak
arises from sampling radio emission from inner cones or core.
Indeed, Gupta \& Gangadhara (2003) find that in all pulsars they have
studied the conal radio emission originates in the inner region
of the open field line cone (with $\rovc$ between $\sim 0.2$ and $\sim
0.8$).
The question of why some pulsars do not exhibit radio peaks
coincident with high-energy peaks is apparently connected to the
question of why some pulsars do not exhibit giant radio pulses.
For three objects (the Crab, B1821$-$24, and B1937$+$21),
all of which have radio peaks coincident with high-energy pulses,
the giant radio pulses have been shown 
to coincide with the high-energy emission
which suggests the same spatial origin in the magnetosphere (Romani \&
Johnston 2001; Cordes et al.~2003; Cusumano et al.~2003).

The polar cap model predicts that the two widely separated peaks of some
gamma-ray pulsars arise when the line of sight crosses the narrow hollow
cone of low altitude gamma ray emission. The nearly aligned geometry
enables one to obtain large peak separation ($\delp \sim 0.4 - 0.5$) in
spite of the narrowness of the gamma-ray beam,
ie.~when $\alpha \sim \theta_\gamma$. Conal radio emission
originates from higher altitudes than the hard gamma-rays, so the nested
radio cones should surround the innermost
gamma-ray cone. Therefore, there should be
radio peaks preceding the leading gamma-ray peak, and following the
trailing gamma-ray peak.
Actually, only one radio peak is usually observed before the leading
gamma-ray peak with no radio peak behind the trailing gamma-ray peak
(the Vela, B1951+52). This can be explained by a commonly observed
lack of trailing parts of radio cones: Lyne \& Manchester (1988)
find that 70 \% of partial conal radio pulsars exhibit 
the lack of the trailing part of radio cones.
However, nearly aligned radio pulsars normally exhibit
very broad radio pulses (Lyne \& Manchester 1988)
which is not the case for the Vela and B1951+52.  
In fact, the width of the radio profile of the Vela pulsar 
is in perfect agreement with a low altitude emission from a rotator with
a large dipole inclination (Rankin 1990). The large $\alpha$ was also
determined for the Vela pulsar via analysis of the Chandra image of its
X-ray nebula (Radhakrishnan \& Deshpande 2001).
The interpretation of the radio peak positions within the polar cap
model is in conflict with these results.

\section{Conclusions}

We find that none of the considered models can account for the observed
variety of both pulsar lightcurves and polarization characteristics.
The two-pole caustic model predicts that the widely separated
double peaks 
observed in the lightcurves of some pulsars
are produced within altitude range between
$\sim 0.2$ and $\sim 0.8\rlc$, where the dipolar structure of the
magnetic field is well justified.
The relative positions of radio and gamma-ray peaks in these
lightcurves
find a natural interpretation within the model.
Interpretation of peaks in the radio pulse of the Crab pulsar is
also possible within the two-pole caustic model. The detailed shape of the 
high-energy lightcurve 
of the Vela pulsar is a clear manifestation of the radiation pattern
predicted generically by the two-pole caustic model
although it must be understood why radio peaks coincident with
high-energy peaks do not appear in
Vela-like pulsars.
However, the model 
is not able to reproduce double peak gamma-ray lightcurves with small peak
separation (like those observed for B1706$-$44 and B1055$-$52,
cf.~Dyks \& Rudak 2003) and the phase lag between the radio and the
gamma-ray peak of B1509$-$58. 
It is also difficult to exactly reproduce
ratios of peak intensity to the bridge, or offpulse intensity
observed for the Crab and Vela pulsars by EGRET.
The polarization characteristics predicted by the caustic model
for some viewing parameters 
(fast swings of the position angle and minima in the
polarization degree at both peaks) resemble those observed in the
optical band for the Crab pulsar. Although
exact agreement cannot be achieved under the
assumptions of this paper, the approximate anticorrelation between 
the received flux and the polarization degree, visible 
in Figs.~\ref{effort1} and \ref{effort2} is worth of notice.
In addition to the Crab pulsar,
the same anticorrelation has been observed in the optical band
for B0656$+$14 (Kern et al.~2003), though the separation of $\sim 0.6$
between the optical peaks of this object cannot be easily interpreted
within the two-pole caustic model.
In the two-pole caustic model, the minima of $\perc[\%]$ associated with
the peaks result naturally from the superposition of radiation originating
from different altitudes -- the same superposition which produces 
the peaks themselves.
This depolarization is efficient only when 
some spatial spread in electron density around the last open magnetic
field lines is assumed.
In addition to this ``caustic depolarization", there are
other minima in $\perc[\%]$ which appear close to the peaks -- these
are caused by the superposition of the two radiation patterns associated
with the opposite magnetic poles.

The position angle curves predicted by the outer gap model
do not resemble the PA curve observed for the Crab.
The outer gap model naturally explains the lightcurves
except for the off-pulse emission, however, it
predicts that the leading gamma-ray peak is produced close to the light
cylinder, and strongly relies on the geometry of magnetic field in this
region. 
To explain the radio properties of 
the Crab pulsar, the outer gap model must locate the radio emission
regions at a variety of positions in the magnetosphere.

The polar cap model requires improbably precise adjustment of viewing geometry
to reproduce observed separations of double peaks in pulsar lightcurves.
It does not predict fast position angle swings at the peaks.

Based on comparison of the predicted and the observed lightcurves, we
find that the two-pole caustic model and the outer gap model
are most successful, although only the two-pole caustic model
is reliable from the point of view of the moderate emission altitudes
involved.
The results described above support
the idea
that the observed high-energy emission from at least some gamma-ray pulsars 
(eg.~Vela, Crab)
has its origin in the outer magnetosphere.
However, the results suggest that the
high-energy
lightcurves of these
pulsars find more natural explanation within the geometry
of the two-pole caustic model rather than within the outer gap model.
The physical basis of the two-pole caustic model may well find its origin in
extended slot gap acceleration (Muslimov \& Harding 2003b).
The discrepancies between the polarization characteristics 
calculated in this paper and
the Crab data do not necessarily disprove the considered models
for three reasons:
1) Our polarization results are determined purely by the geometry of the 
magnetic
field, and do not take into account the frequency dependence
of polarization characteristics (eg.~Epstein 1973; Chen et al.~1996).
2) the Crab pulsar is the \emph{only} case for which the high-energy 
polarization data are available.
3) Polarization characteristics of the pulsed emission from the Crab
pulsar may by affected by the choice of Stokes
parameters for the constant component which is subtracted from the total
signal.

The apparently constant position angle of the constant component 
of Crab optical emission
(Kanbach et al.~2003) is an
astonishing feature 
which may be interpreted in two ways:
1) it may be a contribution from a separate source, not related to the
pulsar magnetosphere (unresolved wisp of the Crab nebula?, some
background object?);
2) it may present emission from a beyond-$\rlc$ region of
the rotating/outflowing dipolar magnetosphere of the
Crab pulsar. 
The second interpretation is possible because in the region beyond the
light cylinder the magnetic field is dominated by a toroidal
component, and the orientation of a toroidal vector 
with respect to the rotation axis is fixed regardless of the rotational
phase (Smith et al.~1988). 
Indeed, Radhakrishnan \& Deshpande (2001) proposed a model for
a Chandra X-ray image of the Vela pulsar, which implies existence
of an apparent, linearly polarized ``jet" aligned with the pulsar rotation axis.
The origin of the ``jet" is the synchrotron radiation from a region
beyond the light cylinder.
The authors note that a similar, as yet spatially
unresolved structure (at X-rays) may
surround the Crab pulsar. Had the synchrotron emission of the ``jet"
extended down to the optical band, it could possibly account for the
observed properties of the constant component of the Crab's optical
emission. However, it would still have to stay spatially 
unresolved even in the optical band.

Assuming the second interpretation is correct, one could explain
the total (pulsed $+$ constant) 
optical emission from the Crab pulsar
as a blend of the pulsed signal predicted by
the two-pole caustic model and the constant emission from the beyond-$\rlc$
region.  
We find that adding a constant component to the theoretical lightcurve 
of the two-pole caustic model does produce the ``zigzag" shape of position angle
curve at the leading peak.

On the other hand, the two component model of optical emission from Crab
(Kellner 2002) is a model only, and it is possible that the total
emission is not a blend of two sources widely separated in space, 
but reflects
the geometry of a single emission region located beyond the light
cylinder. In the latter case extracting the pulsed component
from the total data would be misleading.
Possibly the wind model by Kirk et al.~(2002) offers an example for a source of
radiation which produces fixed position angle at all phases except from 
those at which the peaks occur. 
In the model, high-energy radiation is emitted at large distances from
the star ($10 -100\rlc$) at current sheets, which separate ``layers"
of the toroidal magnetic field. The direction of the magnetic field is
nearly the same at all points within a layer, and it changes by
$180^\circ$ at the current sheets. The sheets are the place where the
magnetic energy is dissipated into particle energy (reconnection), 
and from which the radiation observed at peaks origins.
The pulses are observed because the consecutive sheets
move relativistically towards an observer.
Had the electric field vector of radiation emitted at the sheets
changed its direction by $180^\circ$ (as the direction of $\vec B$ does) 
a fast swing
of position angle would be observed at the peaks. 
Moreover, a disorder in $\vec B$ direction within the reconnection
region could produce
a significant depolarization at the peaks.
Unfortunately, calculations of pulsar polarization properties
have not been performed so far for the wind model.

Finally, let us recall two technical, but interesting results
of this paper: 1) For the case of the magnetic field structure described by
the rotating/retarded dipole, and for moderate inclination angles
$\alpha \simeq 40^\circ-50^\circ$, the polar cap possesses 
a ``notch". At the notch, the magnetic colatitude of the polar cap rim 
is not a single-valued function of magnetic azimuth.
2) Rotational distortions of the magnetic field 
result in low-altitude deformations of polar cap radiation beam
which have magnitude larger than the effects of aberration and
propagation time delays ($\sim \betcor$).

\acknowledgments

We thank G.~Kanbach and F.~Graham-Smith for providing optical
polarization data on the Crab pulsar.
We appreciate fruitful discussions with 
K.~Hirotani, G.~Kanbach, A.~Muslimov, and R.~W.~Romani. 
This work was performed while JD held a National Research
Council Research Associateship Award at NASA/GSFC.
This work was also supported by the grant KBN-2P03D.004.24 (JD and
BR) and by the NASA Astrophysics Theory Program (AH).

\clearpage

\begin{figure}
\epsscale{0.7}
\plotone{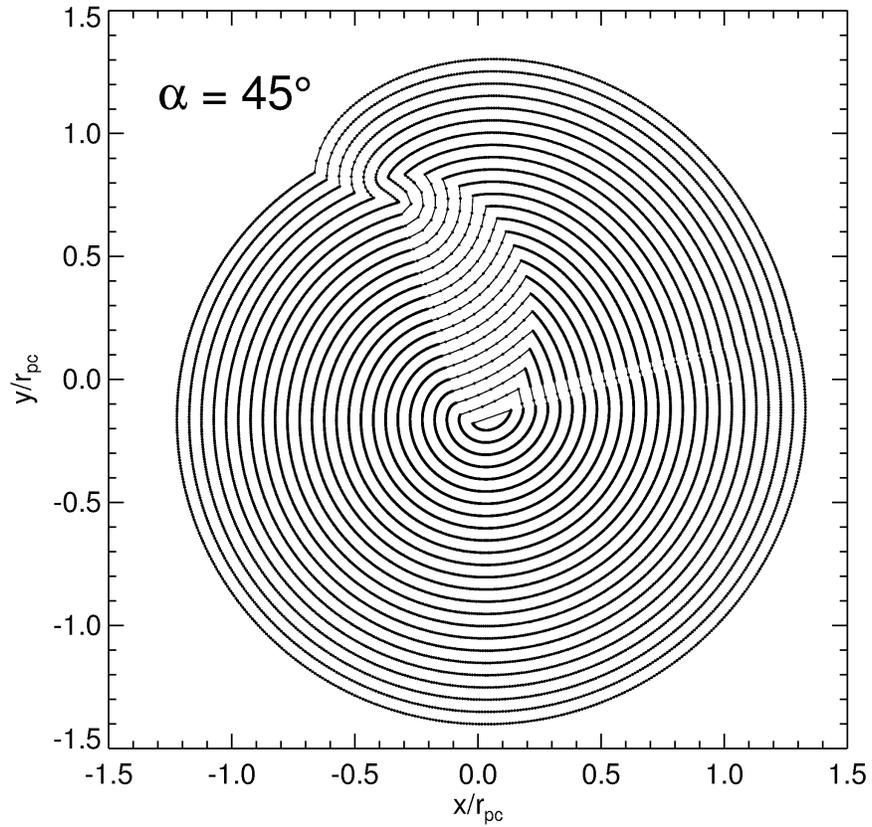}
\caption{The grid of constant open volume coordinate
$\rovc$ for $\alpha = 45^\circ$ and $\prot=0.033$ s. 
The values of $\rovc$ range between $0.05$ (innermost ring) and $1.25$
(outermost) and the distance between them is $0.05$.
The sixth ring (counting inwards) is the rim of the polar cap
(shown with a thick solid line in previous figure).
\label{rings2}}
\end{figure}

\begin{figure}
\epsscale{0.7}
\plotone{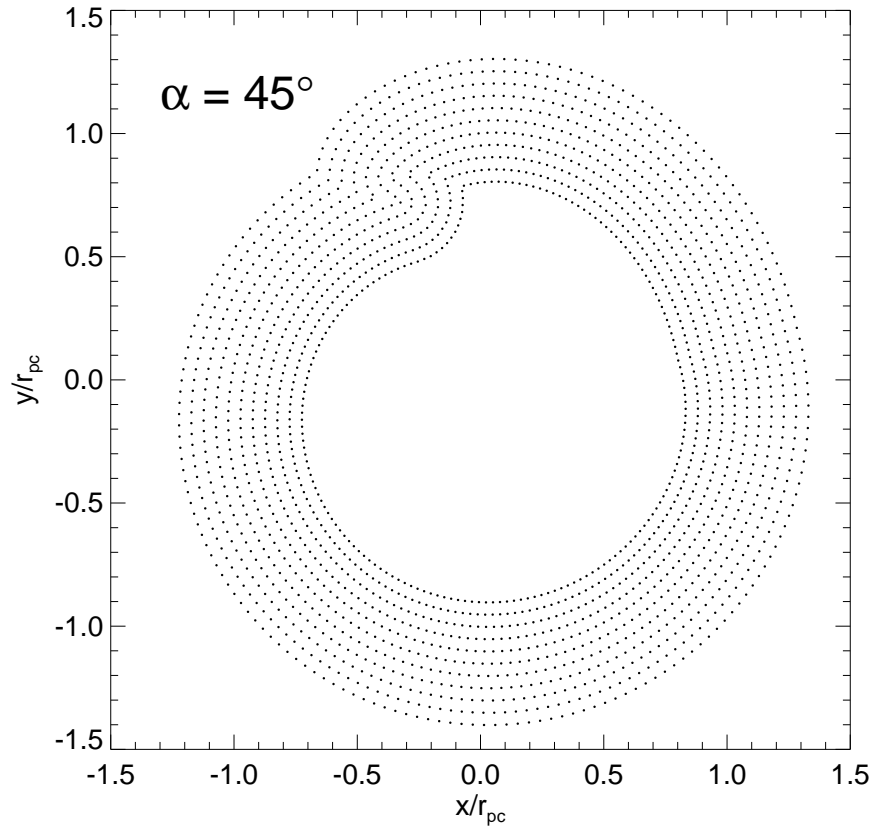}
\caption{Sample distribution of footprints of magnetic field lines
along which photon emission is modelled (for $\alpha = 45^\circ$ and
$\prot=0.033$ s).
There are 180 footprint points uniformly distributed along each ring 
of constant $\rovc$, with the
latter in the range between $0.75$ and $1.25$. The rings are separated
by $0.05$. In calculations described in Section 3 we usually take
a narrower range of $\rovc$ ($0.95 - 1.05$) with smaller ring separation 
of $0.005$, and with 1800 footprints per ring.
\label{inelpos2}}
\end{figure}

\begin{figure}
\epsscale{0.9}
\plotone{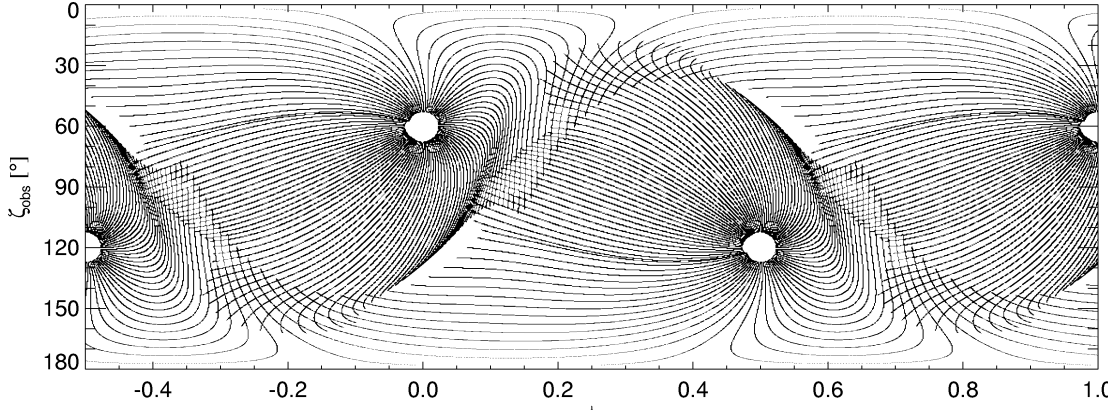}
\caption{Projection of the last open magnetic field lines 
of rotating retarded dipole 
on the space $(\phi,\zobs)$ for $\alpha=60^\circ$. 
Note two strong caustics (dark arches)
trailing both polar caps (blank deformed ovals). They form
because aberration and propagation time effects tend to pile up
at nearly the same phase
photons emitted from different altitudes. 
A horizontal cut of this pattern by an observer 
located at a fixed $\zobs$ produces a lightcurve with two widely
separated peaks. The pattern was calculated for $\rovc = 1$, 
$\rhomax = 0.75\rlc$,
$\rmax = \rlc$, and $\prot=0.033$ s. Only the size of polar caps 
depends on the rotation period $\prot$.  
\label{traj60}}
\end{figure}

\clearpage

\begin{figure}
\epsscale{0.9}
\plotone{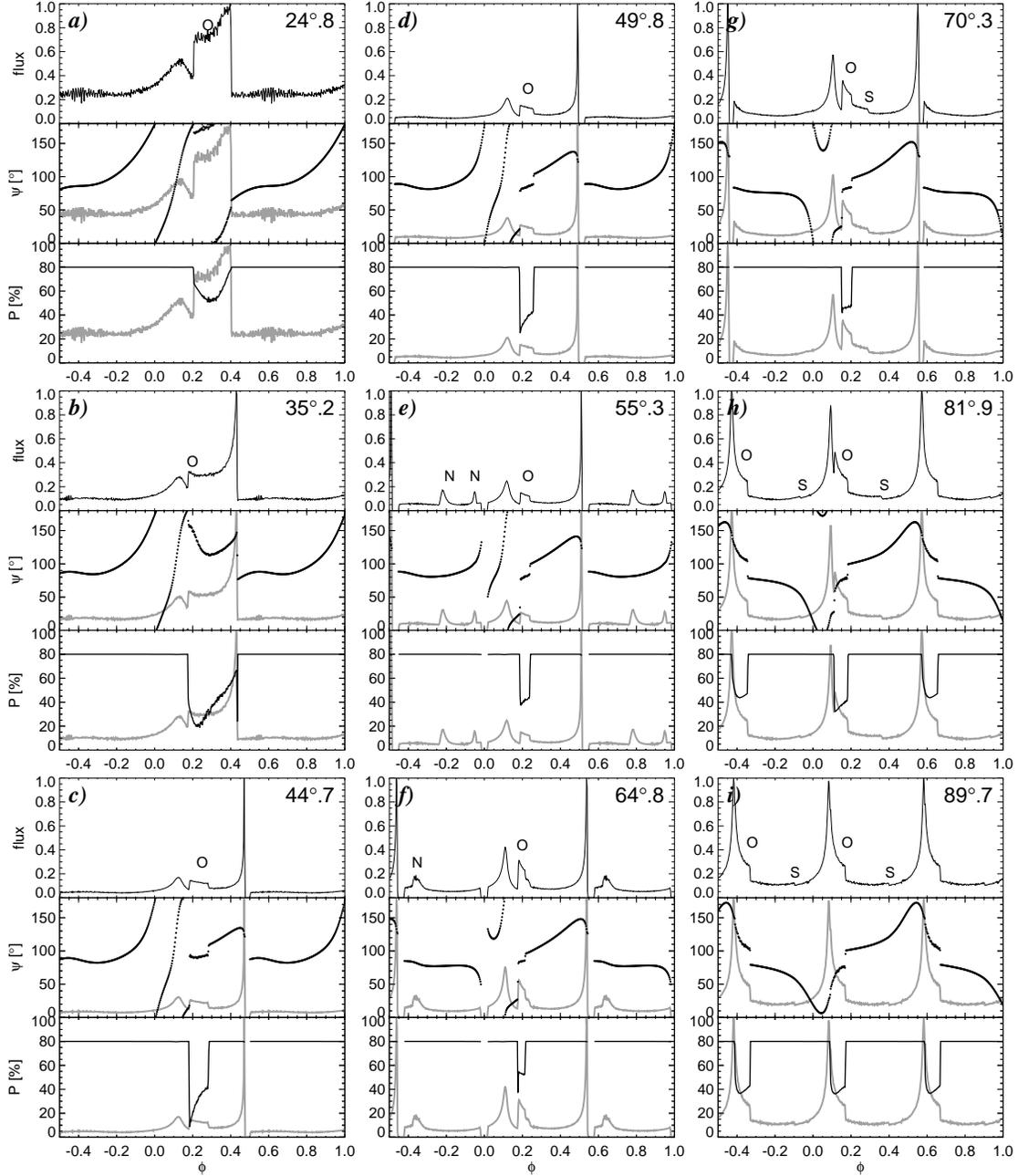}
\caption{Radiation characteristics predicted by the two-pole caustic
model for a pulsar with dipole inclination $\alpha = 60^\circ$.
Nine three-panel frames correspond to nine different viewing angles
$\zobs$ (marked in the top right corners). 
Each frame presents the lightcurve (top panel), the 
position angle curve (dots, middle panel), and the degree of linear
polarization (thick solid line, bottom panel). For reference, 
the lightcurve is overplotted in the middle and in the bottom panels 
as a thick grey line. 
Features marked with N, O, and S are described in the text. 
Note the dominance of two widely separated peaks in lightcurves
for most viewing angles, and a fast swing
of position angle at the first peak for $\zobs \la 70^\circ$.
The results are for  $\rhomax=0.75\rlc$, $\rmax = \rlc$,
$\rovc=1$, and $\prot=0.033$ s. 
\label{rev60}}
\end{figure}

\clearpage

\begin{figure}
\epsscale{0.9}
\plotone{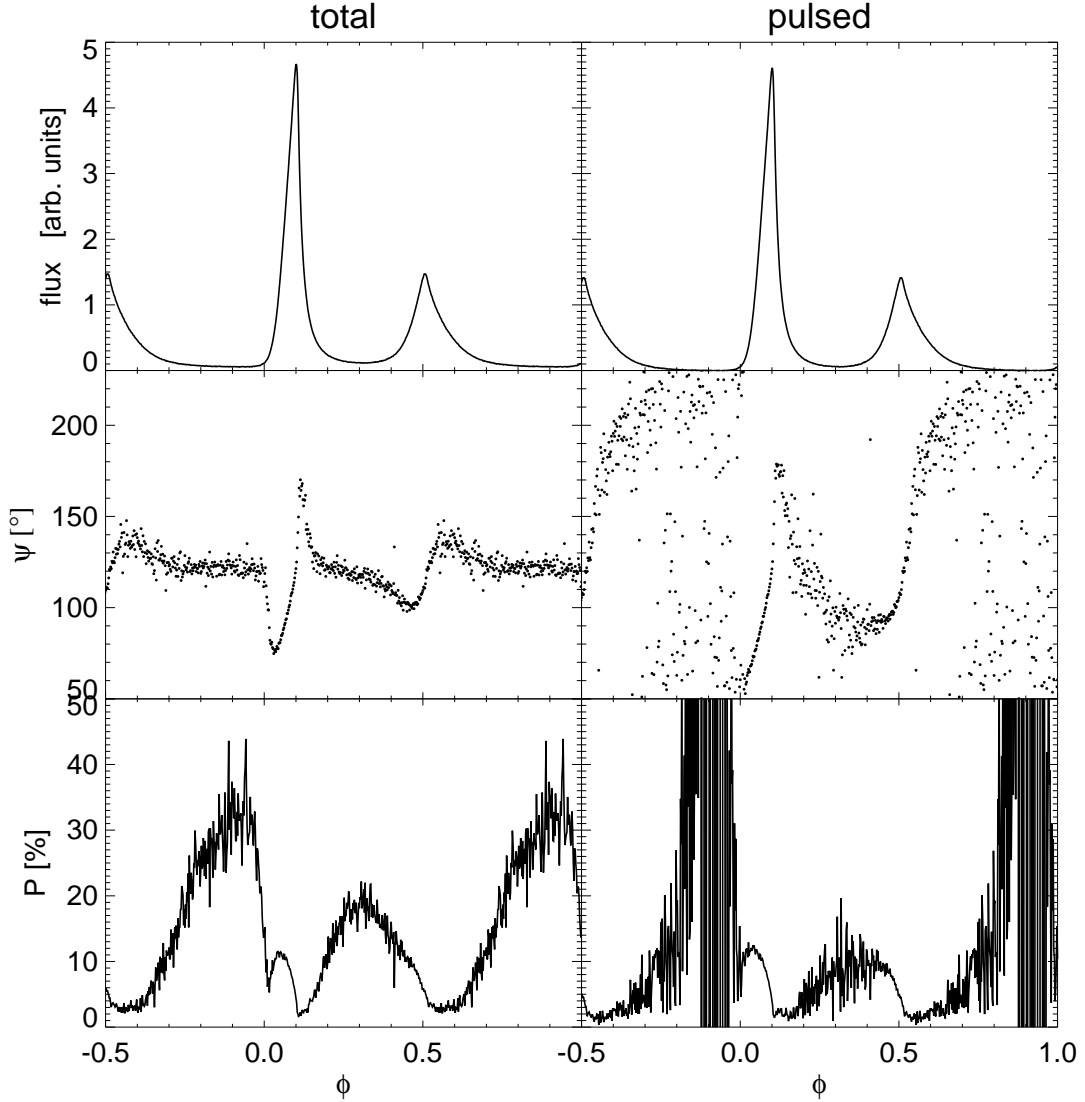}
\caption{Preliminary optical data on the Crab pulsar 
obtained with the OPTIMA instrument (Kanbach et al.~2003).
Left column:
a lightcurve (top panel), a position angle $\psi$ (dots, middle panel),
and a degree of polarization $\perc$ (bottom panel). 
The constant value of position angle within phase ranges $0.6 - 0.9$
and $1.1-1.2$ suggests that the received radiation consists of two
components, one of which has constant properties.
Right column: same as in the left column but with the contribution of
the constant component subtracted from the data. Following Kellner
(2002), for the constant component we assumed intensity equal to $1.24$
\% of the maximum intensity of the total signal, $\psi=123^\circ$,
and $\perc=33$ \%. 
One and a half period is shown. The maximum of the leading peak
was aligned with the phase $\phi=0.1$.
The data were kindly provided by G.~Kanbach.
\label{data}}
\end{figure}

\clearpage

\begin{figure}
\epsscale{0.5}
\plotone{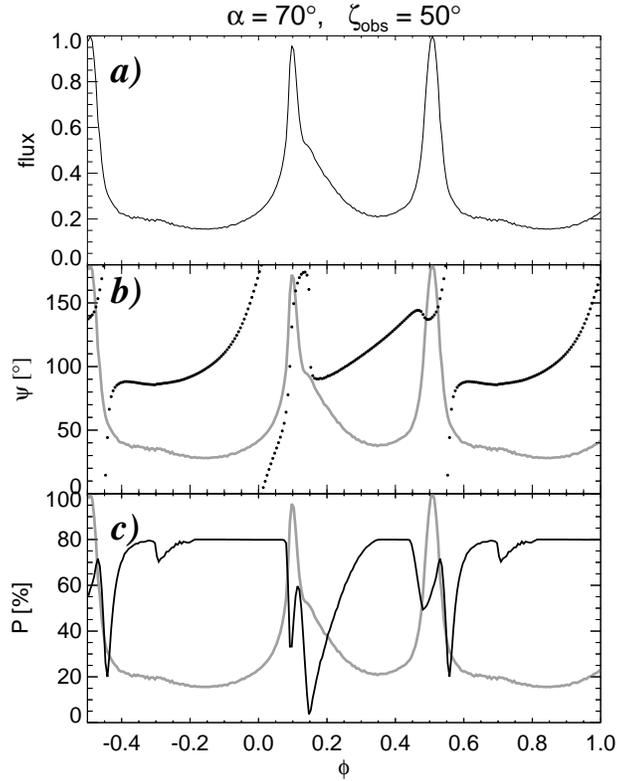}
\caption{A lightcurve (a), a position angle (points, b),
and a degree of polarization (thick solid line, c) 
predicted by the two-pole caustic model
for $\alpha = 70^\circ$
and $\zobs = 50^\circ$.
A spread in electron density at the star surface was assumed
($\sigma = 0.025$, $\rovcz=1$, $\rovcmin = 0.95$, $\rovcmax = 1.05$)
and $\rhomax = 0.8 \rlc$, $\rmax = \rlc$. 
The results are for the retarded dipole field.
\label{effort1}}
\end{figure}

\clearpage

\begin{figure}
\epsscale{0.5}
\plotone{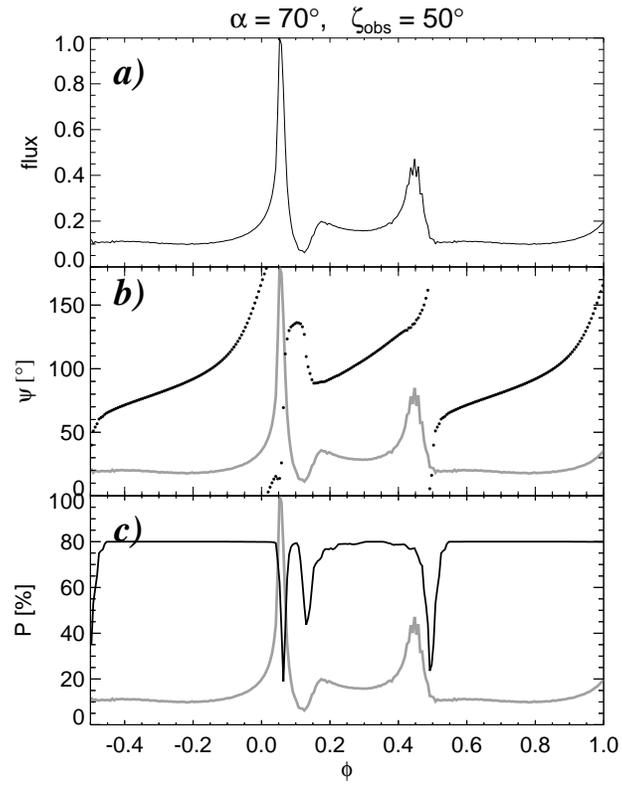}
\caption{Two-pole caustic model results obtained 
for the same parameters 
as in Fig.~\ref{effort1}, but for the static shape
dipole with the circular polar cap rim.
Note the similarity of both these cases.
\label{effort2}}
\end{figure}

\clearpage

\begin{figure}
\epsscale{0.5}
\plotone{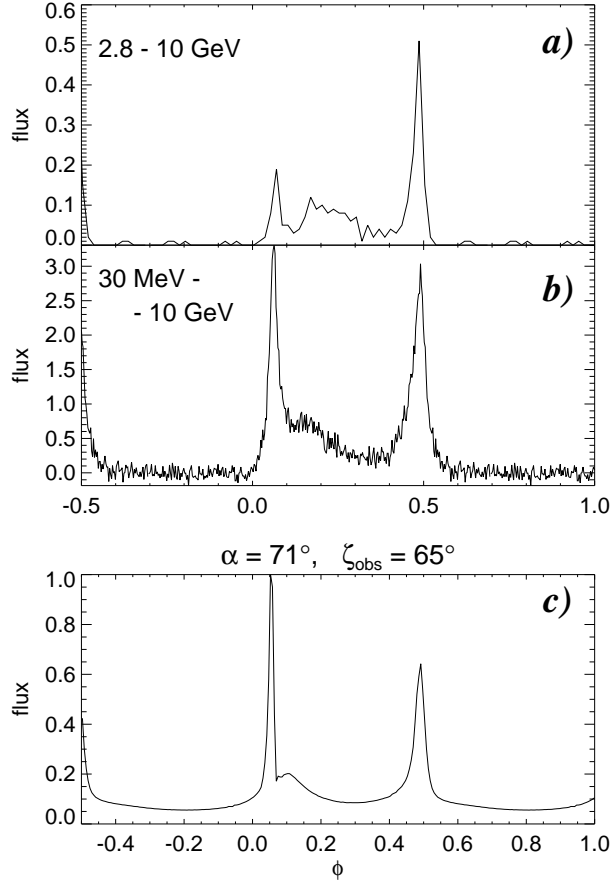}
\caption{Two upper panels present gamma ray lightcurves observed for
the Vela pulsar by EGRET (Kanbach 1999) in two different photon energy
ranges: between $2.8$ and $10$ GeV (a) and between 
$30$ MeV and $10$ GeV (b).
Panel c presents a lightcurve predicted by the two-pole caustic model for
$\alpha=71^\circ$, $\zobs=65^\circ$, $\rmax=\rhomax=0.95\rlc$, and
$\prot=0.0893$ s.
The lightcurve was calculated for
the static shape dipole with the circular polar cap 
and for a spread in the electron density at the star surface with 
$\sigma=0.025$, $\rovcz=1$, $\rovcmin=0.95$ and $\rovcmax=1.05$.
Note the similarity of the model lightcurve (c)
to the one observed within the entire energy band of EGRET
(b). 
\label{effort3}}
\end{figure}

\clearpage

\begin{figure}
\epsscale{0.9}
\plotone{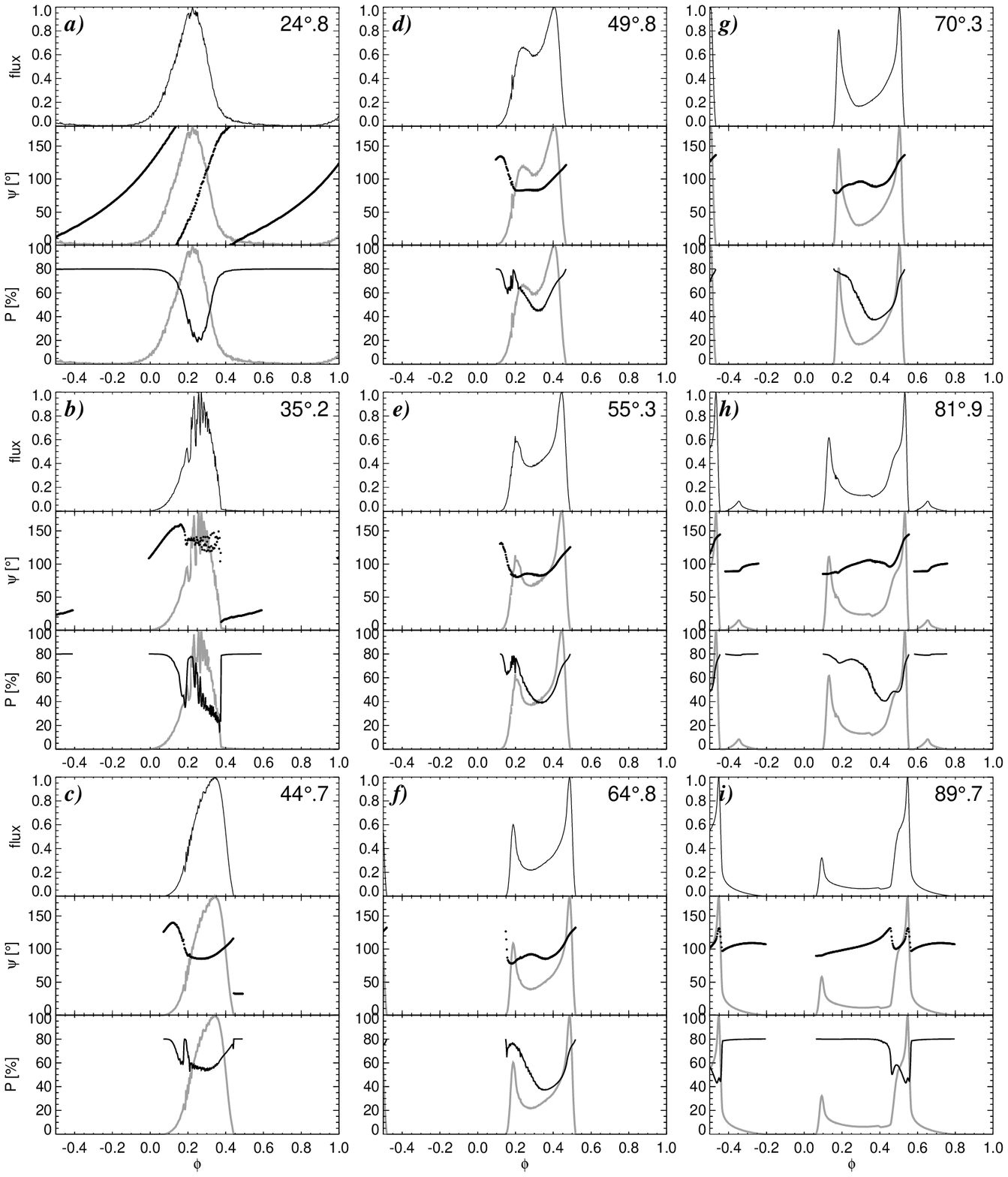}
\caption{Radiation characteristics predicted by the outer gap model for
a pulsar with magnetic dipole inclination $\alpha = 65^\circ$.
The layout is the same as in Fig.~\ref{rev60}.
The results have been obtained for $\rovc=0.9$, $\rhomax = 0.999\rlc$,
$\rmax=1.7\rlc$, $\prot=0.033$ s, and the electron density spread
with $\sigma=0.025$, $\rovcz=0.9$, $\rovcmin=0.85$, and $\rovcmax=0.95$.
\label{revog}}
\end{figure}

\clearpage

\begin{figure}
\epsscale{0.5}
\plotone{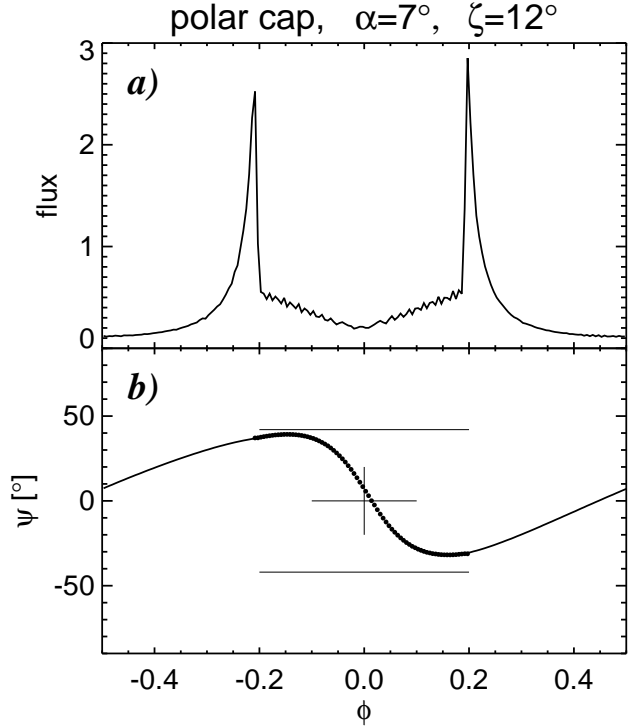}
\caption{Panel a: a gamma-ray lightcurve
predicted by the polar cap model for $\alpha = 7^\circ$,
$\zobs=12^\circ$, $\prot=0.033$ s, and an accelerator at radial distance 
$r=3\rns$.
Panel b: position angle curve. The central, dotted part of the PA curve
($|\phi|\la 0.2$) 
was calculated for photon emission from the
fixed radial distance $r=3\rns$. The rest is for emission from the last open
field lines at $r \ge 3\rns$. The cross centered at $(\phi, \psi) =
(0, 0)$ and the two horizontals at $\psi = \pm 42^\circ$ provide
reference to discern the influence of rotation on the PA curve
(see the text for details). The results are for the static shape dipole
with the circular polar cap.}
\label{polcap}
\end{figure}

\clearpage

\begin{figure}
\epsscale{0.7}
\plotone{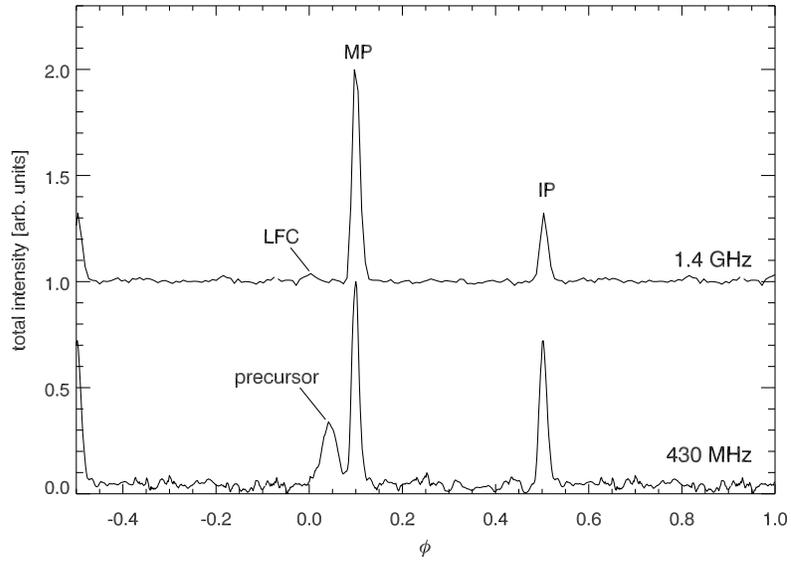}
\caption{Radio pulse profiles of the Crab pulsar at $1.4$ GHz (upper
curve) and $430$ MHz (lower curve). The pulse features discussed in
Section \ref{altitudes} are identified. One and a half period is shown.
The main peak (MP) of the profiles was aligned with the phase
$\phi=0.1$. The data are from Moffett \& Hankins (1996).}
\label{radiocrab}
\end{figure}

\clearpage

\begin{figure}
\epsscale{0.5}
\plotone{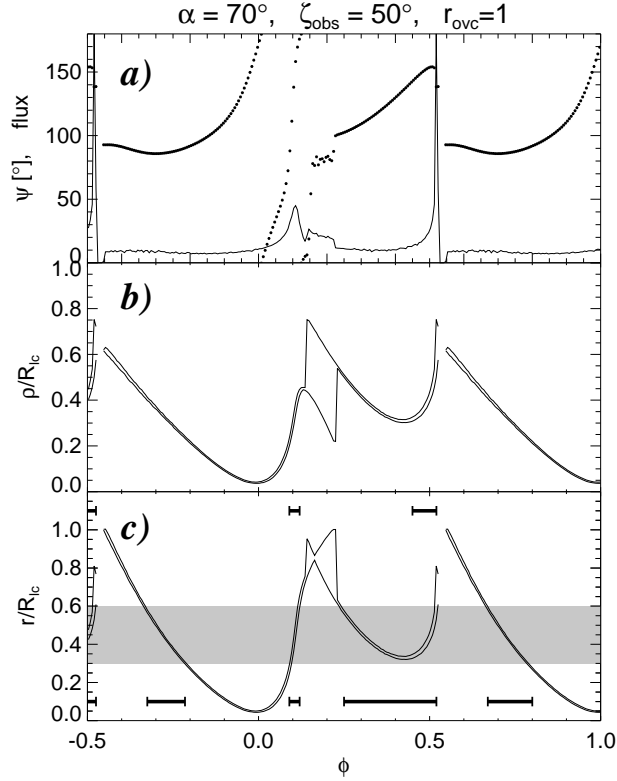}
\caption{Panel a: a lightcurve (solid line), and a position angle curve
(dots) predicted by the two-pole caustic model for $\alpha=70^\circ$,
$\zobs=50^\circ$, $\rovc=1$, $\rhomax=0.75\rlc$ and $\rmax=\rlc$.
Panel b: the distance $\rho$ of emission points from the rotation axis 
as a function of phase $\phi$ at which the radiation emitted at these points
is detected.
The bottom curve is for the minimum value of the distance, the top curve
is for the maximum distance.
Panel c: the radial distance $r$ of emission points from the star center
as a function of the detection phase $\phi$.
The bottom curve is for the minimum value of $r$, 
the top line -- for the maximum $r$.
The shaded band denotes the range of $r$ from which coherent radio
waves should be emitted 
in order for radio peaks
to coincide with the gamma-ray peaks visible in
panel a. Projection of those fragments of the $r(\phi)$ curves which
cross the shaded band onto the horizontal axis determines ranges of
phase within which radio emission would be observed (marked with
horizontal bars near the bottom horizontal axis). In addition to the
radio peaks at $\phi=0.1$ and $\phi=0.5$, a bridge, and an offpulse
radio emission would be observed. If only a trailing part of
radio emission cones existed, the radio emission from the outermost
cones
would be observed only at phases coincident with the gamma ray peaks
(horizontal bars near the top horizontal axis).
Radio emission from inner cones could produce the LFC and the radio precursor.
\label{radii1}}
\end{figure}

\clearpage

\begin{figure}
\epsscale{0.5}
\plotone{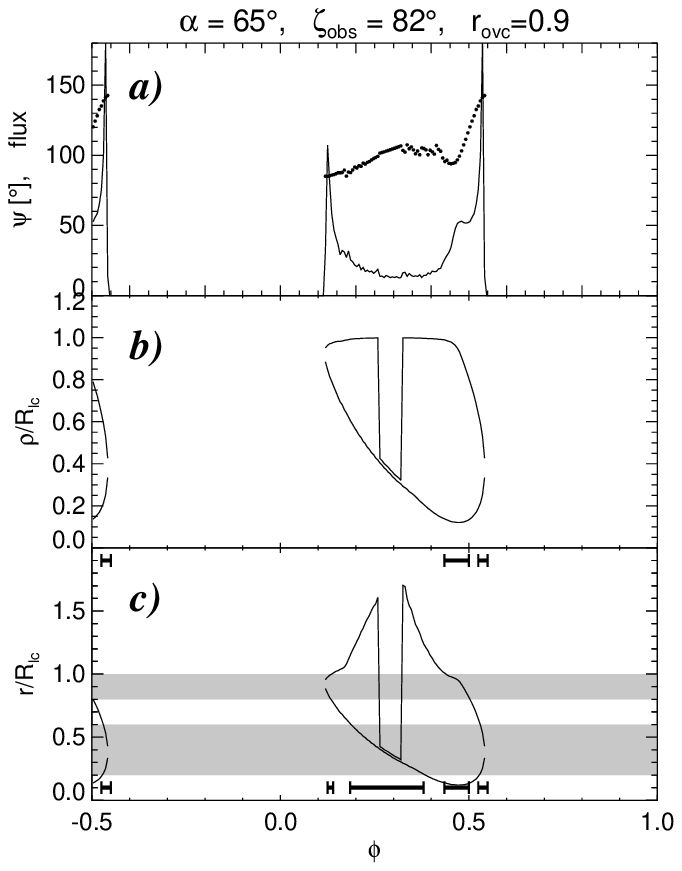}
\caption{The same as in Fig.~\ref{radii1} but for the outer gap model
with $\alpha = 65^\circ$, $\zobs=82^\circ$, $\rovc=0.9$,
$\rhomax=0.999\rlc$, and $\rmax = 1.7\rlc$ (the same parameters as in 
Fig.~7 of CRZ2000).
\label{radii2}}
\end{figure}


\begin{thebibliography}{}
\bibitem[1998]{ae98} Arendt, P. N., \& Eilek, J. A. 1998, \apj,
  submitted (astro-ph/9801257)
\bibitem[1983]{arons} Arons, J. 1983, \apj, 266, 215
\bibitem[1979]{as79} Arons, J., \& Scharlemann E. T. 1979, \apj, 231,
  854
\bibitem[2001]{b2001} Baring, M. G.
  2001, in Astrophysics and Space Science Library 267,
  Proc. Tonantzintla Workshop, ed. 
  A. Carrami{\~ n}ana et al. (Dordrecht), 167
\bibitem[1991]{blask91} Blaskiewicz, M., Cordes, J. M., \& Wasserman, I.
  1991, \apj, 370, 643 (BCW91)
\bibitem[2003]{peter} Bloser, P. F., Hunter, S. D., Depaola, G. O.,
  \& Longo, F. 2003, in Proc.~SPIE 5165, ``X-ray and Gamma-Ray
  Instrumentation for Astronomy XIII", in press
\bibitem[1996]{chen96} Chen, K., Chang, H.-K., \& Ho, C. 1996, \apj,
  471, 967
\bibitem[1986]{chr86} Cheng, K. S., Ho, C., \& Ruderman, M. 1986, \apj,
  300, 500
\bibitem[2000]{crz00} Cheng, K. S., Ruderman, M. A., \& Zhang, L. 2000,
  \apj, 537, 964 (CRZ2000)
\bibitem[2003]{cordes2003} Cordes, J. M., Bhat, N. D. R., Hankins, T.
  H., McLaughlin, M. A., \& Kern, J. 2003, \apj, submitted 
  (astro-ph/0304495)
\bibitem[2003]{ck2003} Crawford, F., \& Keim, N. C. 2003, \apj, 590,
  1020
\bibitem[2001]{cmk2001} Crawford, F., Manchester, R. N., \&
  Kaspi, V. M. 1991, \aj, 122, 2001
\bibitem[2003] {cusumano2003} Cusumano, G., Hermsen, W., Kramer, M., 
  Kuiper, L., L\"ohmer, O., et al. 2003, A\&A, submitted
  (astro-ph/0309580)
\bibitem[1992]{dt92} Damour, T., \& Taylor, J. H. 1992, Phys. Rev. D,
  45, 1840 
\bibitem[1982]{dau82} Daugherty, J.K., \& Harding, A.K. 1982, ApJ,  252,
  337
\bibitem[1994]{dau94} Daugherty, J.K., \& Harding, A.K. 1994, ApJ,  42,
  325
\bibitem[1996]{dau96} Daugherty, J.K., \& Harding, A.K. 1996, ApJ, 458,
  278
\bibitem[2002]{dyks} Dyks, J. 2002, PhD Thesis, Nicolaus Copernicus
    Astronomical Center
\bibitem[2000]{dr2000} Dyks, J., \& Rudak, B. 2000, \mnras, 319, 477
\bibitem[2002]{dr2002} Dyks, J., \& Rudak, B. 2002, A\&A, 393, 511
\bibitem[2003]{dr2003} Dyks, J., \& Rudak, B. 2003, \apj, 598, 1201
\bibitem[1973]{e73} Epstein, R. I. 1973, \apj, 183, 593
\bibitem[2001]{ew2001} Everett, J. E., \& Weisberg, J. M. 2001, ApJ,
  553, 341 (EW2001)
\bibitem[1995]{fier95} Fierro, J. M. 1996, PhD Thesis, Stanford
  University
\bibitem[2001]{gg2001} Gangadhara, R. T., \& Gupta, Y. 2001, \apj, 555,
  31
\bibitem[1998]{gkm98} Gil, J., Khechinashvili, D. G., \& Melikidze, G.
 I. 1998, \mnras, 298, 1207
\bibitem[1996]{gk96} Gil, J., \& Krawczyk, A. 1996, \mnras, 280, 143
\bibitem[1996]{gs1996} Graham-Smith, F., Dolan, J. F., Boyd, P. T., 
  Biggs, J. D., Lyne, A. G., et al. 1996, \mnras, 282, 1354
\bibitem[1988]{ghc} Grenier, I. A., Hermsen, A., \& Clear, J. 1988, A\&A,
    204, 117
\bibitem[2003]{gg2003} Gupta, Y., \& Gangadhara, R. T. 2003, \apj, 584,
    418
\bibitem[1978]{hte} Harding, A. K., Tademaru, E., \& Esposito, L. W.
    1978, \apj, 225, 226
\bibitem[2001]{hel2001} Helfand, D. J., Gotthelf, E. V., \& Halpern, J.
  P. 2001, \apj, 556, 380
\bibitem[1995]{hest95} Hester, J. J., Scowen, P. A., Sankrit, R., 
  Burrows, C. J., Gallagher III, J. S., et al. 1995, \apj, 448, 240
\bibitem[2001]{ha2001} Hibschman, J. A., \& Arons, J. 2001, ApJ, 546,
  382 (HA2001)
\bibitem[2001]{hs2001} Hirotani, K., \& Shibata, S., 2001, \mnras, 325, 1228
\bibitem[2003]{hs2003} Hirotani, K., Harding, A.K., \& Shibata, S.
  2003, \apj, 591, 334
\bibitem[2001]{joh2001} Johnston, S., van Straten, W., Kramer, M., \&
  Bailes, M. 2001, \apj, 549, L101
\bibitem[1981]{jon81} Jones, D. H. P., Smith, F. G., \& Wallace, P. T.
  1981, \mnras, 196, 943
\bibitem[1994]{kab} Kanbach, G., Arzoumanian, Z., Bertsch, D. L., 
 Brazier, K. T. S., Chiang, J., et al. 1994, A\&A, 289, 855
\bibitem[1999]{kan99} Kanbach, G. 1999, 
 in Astroph. Lett. Comm. 38, Proc.~of the Third INTEGRAL Workshop, 17
\bibitem[2003]{kan2003} Kanbach, G., Kellner, S., Schrey, F., Steinle,
  H., Straubmeier, C., et al. 2003, SPIE Meeting on Astronomical
  Telescopes and Instrumentation ``Power Telescopes and Instrumentation
  into the New Millenium", eds.~Iye, M., \& Moorwood, A. F., 4841, 82
\bibitem[2002]{kel2002} Kellner, S. 2002, Diplomarbeit, Technische
  Universit\"at M\"unchen
(http://www.mpe.mpg.de/gamma/instruments/optima/www/optima-papers.html)
\bibitem[2003]{kern2003} Kern, B., Martin, C., Mazin, B., \& Halpern,
  J. P. 2003, ApJ, accepted (ApJ preprint doi:10.1086/378670)
\bibitem[2001]{kij2001} Kijak, J. 2001, \mnras, 323, 537
\bibitem[2002]{kg2002} Kijak, J., \& Gil, J. 2002, A\&A, 392, 189
\bibitem[2002]{kirk} Kirk, J. G., Skjaeraasen, O., \& Gallant, Y. A.
  2002, A\&A, 388, L29
\bibitem[1983]{kd83} Krishnamohan, S., \& Downs, G. S. 1983, \apj, 265,
  372
\bibitem[2001]{kuiper2001} Kuiper, L., Hermsen, W., Cusumano, G., 
  Diehl, R., Sch\"onfelder, V., et al.~2001, A\&A, 378, 918
\bibitem[2003]{kuiper2003} Kuiper, L., Hermsen, W., Walter, R., \&
  Foschini, L. 2003, A\&A, submitted (astro-ph/0309178)
\bibitem[2001]{lai2001} Lai, D., Chernoff, D. F., \& Cordes, J. M. 2001,
  ApJ, 549, 1111
\bibitem[1988]{lm88} Lyne, A. G., \& Manchester, R. N. 1988, \mnras,
  234, 477
\bibitem[1998]{malo} Malofeev, V. M. 1998, Astron.~Zh., 75, 281
\bibitem[2000]{mm} Malofeev, V. M., \& Malov, O. I. 2000, Astron.~Zh.,
  77, 52
\bibitem[1979]{smb} Massaro, E., Salvati, M., \& Buccheri, R. 1979, \mnras,
    189, 823
\bibitem[1996]{mf96} Moffett, D. A., \& Hankins, T. H. 1996, \apj, 468,
  779
\bibitem[1999]{mf99} Moffett, D. A., \& Hankins, T. H. 1999, \apj, 522,
  1046
\bibitem[1983]{mor83} Morini, M. 1983, \mnras, 202, 495 
\bibitem[2003a] {mh2003a} Muslimov, A. G., \& Harding, A. K. 2003a, \apj, 
   588, 430
\bibitem[2003b] {mh2003b} Muslimov, A. G., \& Harding, A. K. 2003b, \apj, 
   submitted
\bibitem[2000]{pav2000} Pavlov, G. G., Sanval, D., Garmire, G. P.,
  Zavlin, V. E., Burwitz, V., et al. 2000, AAS Meeting, 196, 37.04 
\bibitem[1969]{rc1969} Radhakrishnan, V., \& Cooke, D. J. 1969,  
  Astrophys. Lett., 3, 225
\bibitem[2001]{rd2001} Radhakrishnan, V., \& Deshpande, A. A. 2001, 
 A\&A, 379, 551
\bibitem[1990]{ran90} Rankin, J. M. 1990, \apj, 352, 247
\bibitem[1993]{ran93} Rankin, J. M. 1993, \apj, 405, 285
\bibitem[2001]{rj2001} Romani, R. W., \& Johnston, S. 2001, \apj, 590,
  L95
\bibitem[2001]{rom2001} Romani, R. W., Miller, A. J., Cabrera, B., 
  Nam, S. W., \& Martinis, J. M. 2001, \apj, 563, 221
\bibitem[1995]{ry1995} Romani, R. W. \& Yadigaroglu, I.-A., 1995, \apj,
438, 314 (RY95)
\bibitem[2000]{rots2000} Rots, A. H., Jahoda, K., \& Lyne, A. G. 2000,
  AAS HEAD Meeting, 32, 3308
\bibitem[1999]{rd99} Rudak, B., \& Dyks, J. 1999, \mnras, 303, 477
\bibitem[1975]{rs75} Ruderman, M. A., \& Sutherland, P. G. 1975, \apj,
196, 51
\bibitem[1988]{smith88} Smith, F. G., Jones, D. H. P., Dick, J. S. B., 
  \& Pike, C. D. 1988, \mnras, 233, 305
\bibitem[1995]{sdm95} Sturner, S. J., Dermer, C. D., \& Michel, F. C.
1995, \apj, 445, 736
\bibitem[1971]{st71} Sturrock, P. A. 1971, \apj, 164, 529
\bibitem[2001]{tennant2001} Tennant, A. F., Becker, W., Juda, M., 
  Elsner, R. F., Kolodziejczak, J. J., et
  al.~2001, ApJ, 554, L173
\bibitem[2001]{thomp2001} Thompson, D. J. 2001, in AIP Proceedings 558,
 High Energy Gamma-Ray Astronomy, ed. A. Goldwurm et al., 103
\bibitem[2003]{w2003} Wright, G. A. E. 2003, \mnras, 344, 1041
\bibitem[1995]{y97} Yadigaroglu, I.-A. 1997, Ph.D. thesis, Stanford
  University
\end{thebibliography}
\end{document}